%% file: main.tex
\documentclass[journal]{vgtc}        

\onlineid{1332}

\vgtccategory{Research}

\vgtcpapertype{evaluation}

\title{VisEval: A Benchmark for Data Visualization in the Era of Large Language Models}

\author{%
    Nan Chen, Yuge Zhang, Jiahang Xu, Kan Ren, and Yuqing Yang
}

\authorfooter{
  \item
    Nan Chen, Yuge Zhang, Jiahang Xu, and Yuqing Yang are with Microsoft Research while Kan Ren is with ShanghaiTech University. Yuqing Yang and Kan Ren are corresponding authors.
    
    E-mails: \{nanchen,Yuge.Zhang,jiahangxu,yuqing.yang\}@microsoft.com. renkan@shanghaitech.edu.cn.
}

\abstract{
    Translating natural language to visualization (NL2VIS) has shown great promise for visual data analysis, but it remains a challenging task that requires multiple low-level implementations, such as natural language processing and visualization design.
    Recent advancements in pre-trained large language models (LLMs) are opening new avenues for generating visualizations from natural language.
    However, the lack of a comprehensive and reliable benchmark hinders our understanding of LLMs' capabilities in visualization generation.
    In this paper, we address this gap by proposing a new NL2VIS benchmark called \name.
    Firstly, we introduce a high-quality and large-scale dataset. This dataset includes 2,524 representative queries covering 146 databases, paired with accurately labeled ground truths.
    Secondly, we advocate for a comprehensive automated evaluation methodology covering multiple dimensions, including validity, legality, and readability.
    By systematically scanning for potential issues with a number of heterogeneous checkers, \name provides reliable and trustworthy evaluation outcomes.  
    We run \name on a series of state-of-the-art LLMs. Our evaluation reveals prevalent challenges and delivers essential insights for future advancements.
}


\teaser{
  \centering
  \includegraphics[width=\linewidth, alt={A view of a city with buildings peeking out of the clouds.}]{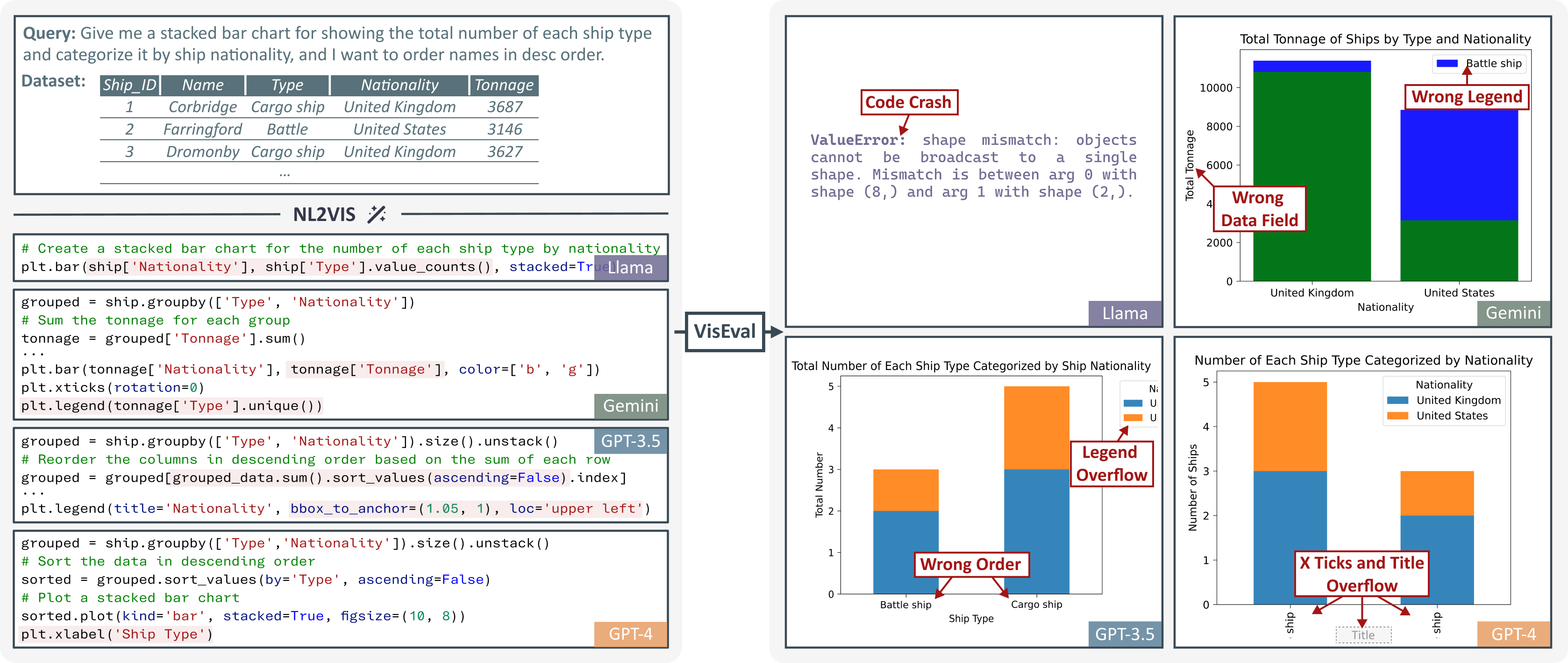}
  \caption{Examples of generating visualization using LLMs. Llama (CodeLlama-7B) produces code that cannot be executed. Gemini (Gemini-Pro) incorrectly maps the ``sum of Tonnage'' to the y-axis instead of ``count'' and lacks a legend for the ``Cargo ship'' color. GPT-3.5 fails to sort as specified and positions the legend outside the canvas. Although GPT-4 almost meets the requirement, it still encounters overflow issues that impact readability.}
  \label{fig:example}
}

\graphicspath{{figs/}{figures/}{pictures/}{images/}{./}} 

\usepackage{mathptmx}                  

\usepackage{xspace}
\usepackage{graphicx}
\usepackage{float}
\usepackage[dvipsnames, svgnames]{xcolor}
\usepackage{color}
\usepackage{colortbl}
\usepackage{multirow}
\usepackage{array}
\usepackage{threeparttable}
\usepackage{multicol}
\usepackage{booktabs}
\usepackage{tikz}
\usepackage{soul}
\usepackage{rotating} 
\usepackage{graphicx} 
\usepackage{amssymb}

\usepackage[title]{appendix}

\definecolor{legend}{RGB}{191,189,55}
\definecolor{minus}{RGB}{46,146,58}

\newcommand{\etal}{{\it et~al.}\xspace}
\newcommand{\name}{VisEval\xspace}

\newcommand{\revise}[1]{{#1}}

\newcommand{\error}[1]{\textcolor{orange}{#1}}
\newcommand{\minus}[1]{\textcolor{minus}{#1}}
\newcommand{\red}[1]{\fontsize{6.0pt}{\baselineskip}\selectfont\textcolor{red}{#1}}

\begin{document}



\input {sections/01-intro}

\input {sections/02-related}

\input{sections/03-preliminaries}

\input{sections/04-methodology}

\input{sections/05-evaluation}

\input{sections/06-discussion}

\input{sections/07-conclusion}

\bibliographystyle{abbrv-doi-hyperref}

\bibliography{main}

\appendix 

\clearpage

\input{sections/appendix}

\end{document}

%% file: sections/01-intro.tex
\firstsection{Introduction}

\maketitle

NL2VIS, the task of translating natural language (NL) queries based on provided data tables into visualizations (VIS), has been a longstanding goal in the field of data visualization~\cite{narechania2020nl4dv,shen2022towards,zhang2023natural,he2024leveraging}. 
It bridges the gap between human understanding and complex data, enabling users to handle intricate data analysis or visualization requirements in a user-friendly manner.
The challenges of NL2VIS were multifaceted, with difficulties ranging from accurately interpreting NL queries to effectively transforming data and selecting meaningful visual mappings~\cite{narechania2020nl4dv,shen2022towards}. 
For instance, query interpretation involves grappling with the intricacies of natural language, while data transformation necessitates handling diverse data sources and formats. Additionally, visual mapping needs to satisfy the diverse demands of visualization.

Recently, pre-trained large language models (LLMs)~\cite{touvron2023llama,openai2023gpt4} have demonstrated outstanding performance across various natural language-related tasks, such as data science~\cite{zhang2024benchmarking}, code generation~\cite{chen2021evaluating}, and web design~\cite{kim2022stylette}. 
This success brings hope for addressing the challenges mentioned above. 
LLMs-based methods have rapidly emerged as the predominant approach for addressing NL2VIS tasks.
For instance, Chat2vis~\cite{maddigan2023chat2vis} and LIDA~\cite{dibia2023lida} have demonstrated proficiency in generating data visualizations through prompt tuning or engineering. 
Moreover, ChartLlama~\cite{han2023chartllama} and ChartGPT~\cite{tian2024chartgpt} leverage the training or fine-tuning of LLMs to develop specialized models for visualization, thereby further enhancing their capabilities in solving NL2VIS tasks.

Without loss of generality, the typical workflow of visualization generation using LLMs entails assembling an NL query and serialized data tables into a prompt, then soliciting LLMs to generate code based on an established visualization library (e.g., Matplotlib~\cite{hunter2007matplotlib}, Vega-Lite~\cite{satyanarayan2016vega}).
The code is then executed in a sandboxed environment to obtain the final chart.
Regrettably, this process occasionally encounters errors, leading to flawed outcomes.
As illustrated in~\cref{fig:example}, when we visualize the ship dataset with a stacked bar chart, state-of-the-art LLMs all suffer from various issues.
These issues range from code execution failures to incorrect data transformations, illegal sorting, and text extending beyond the canvas.
Visualizations may appear correct at first glance, but they contain easily overlooked issues that can mislead users~\cite{hopkins2020visualint,gu2023analysts,lo2022misinformed}.
Such shortcomings highlight the urgent need for systematic evaluation and benchmarking that points out potential issues in the generated results and reporting reliable evaluation outcomes.



However, current practices of NL2VIS evaluations fall short of adequately addressing this need, due to limitations in the quality and scalability of datasets, the comprehensiveness of metrics, and the reliability of methodologies.
Mainstream NL2VIS datasets either focus on narrow domains and lack scalability~\cite{gao2015datatone, kumar2016towards, srinivasan2021collecting}, or contain incorrect labels and ambiguous queries~\cite{luo2021synthesizing}.
The comprehensiveness of evaluation is also a long-existing issue. For example, some evaluations~\cite{maddigan2023chat2vis1, li2024visualization} solely look at the correctness of presented data, neglecting other dimensions such as readability.
Some other studies~\cite{dibia2023lida, han2023chartllama} take various metrics into consideration and leverage LLMs to assess the code generated by themselves, but LLMs-powered evaluations remain inadequately scrutinized in terms of proficiency, leading to doubts about the reliability. 
To the best of our knowledge, no existing benchmarks contain both high-quality and large-scale datasets along with reliable automated evaluation methodologies that cover diverse metrics.

To fill this gap, we introduce \name, a novel NL2VIS benchmark that thoroughly and reliably evaluates generated visualizations.
We start by constructing a dataset, comprising 2,524 representative natural language queries covering 146 databases. 
Aiming to create a dataset with large-scale coverage, high-quality queries, accurate ground truth, and valuable selected queries, we implement a data filtering procedure that combines the intelligence of state-of-the-art LLMs and experiences from visualization experts.
We also introduce a novel labeling procedure that annotates meta-information that defines the feasible region for multiple acceptable \revise{charts}, rather than the exact match of a single one~\cite{maddigan2023chat2vis1}. 
\revise{Finally, we rebalance the dataset to a moderate difficulty.}

Next, an automated evaluation framework is designed to comprehensively scan for issues related to validity, legality, and readability.
As shown in~\cref{fig:evaluator}, the validity checker executes the code and verifies its capability to generate visualizations, thereby ensuring the \textbf{validity}.
Following that, the legality checker deconstructs the visualization to extract information such as chart type and data. A series of checks then impartially examine the \textbf{legality} of chart type, data, and sorting with the aid of annotated meta-information from the dataset.
Finally, the assessment of \textbf{readability} is the most challenging and complex part, requiring consideration of various factors such as layout, scale, and color, which makes it hard to achieve through predefined rules.
We leverage the power of GPT-4V(ISION)~\cite{openai2023gpt4v} and implement an automated workflow to evaluate readability. 
Our quantitative experiments show that the readability evaluator is well-aligned with human preferences.

Based on the constructed dataset and the well-designed evaluation framework, we conducted a comprehensive evaluation of state-of-the-art LLMs, including GPT-4~\cite{openai2023gpt4}, GPT-3.5~\cite{ouyang2022training}, Gemini-Pro~\cite{team2023gemini}, and CodeLlama-7B~\cite{roziere2023code}.
The results of evaluations reveal the common challenges and limitations and provide useful insights for future advancements.
To summarize, our contributions are as follows:
\begin{itemize}[noitemsep,topsep=0pt,leftmargin=10pt] 
    \item We construct a high-quality and large-scale dataset with accurate ground truth, supplemented by meta-information for evaluation.
    \item We introduce a novel and reliable evaluation framework for a comprehensive assessment of the generated visualizations, covering various dimensions including validity, legality, and readability.
    \item We conduct comprehensive evaluations of state-of-the-art LLMs from various perspectives, shedding light on their capability and unveiling avenues for advancement.
\end{itemize}


%% file: sections/02-related.tex
\section{Related Work}


\subsection{Natural Language to Visualization Generation}
Over the years, natural language has proven to be an efficient way of specifying visualization~\cite{kavaz2023chatbot,zhang2023natural}.
Traditional methods ~\cite{gao2015datatone,yu2019flowsense,narechania2020nl4dv} utilized semantic or lexical parsing techniques to infer user intent and then return appropriate visualizations.
Recently, deep learning-based methods have further advanced the development of methods~\cite{luo2021natural,song2022rgvisnet} for translating natural language into visualizations.
Despite notable enhancements achieved in NL2VIS task, limitations persist in generalization, primarily due to constraints in predefined rules or datasets.

The emergence of LLMs introduces a new direction for data visualization generation, exhibiting excellent generalization capabilities.
In most studies~\cite{maddigan2023chat2vis,dibia2023lida,hong2023conversational,li2024prompt4vis}, LLMs are used to directly generate visualization through prompt tuning and engineering. 
For example, Chat2vis~\cite{maddigan2023chat2vis} utilizes prompts containing natural language queries and a textual description of tabular data, incorporating column names and values, for generating Python visualization code using LLMs.
Similarly, LIDA~\cite{dibia2023lida} defines visualization generation as a four-stage problem and generates visualization based on established Python visualization libraries.
Another branch of research involves training or fine-tuning LLMs to develop models specialized for visualization.
For example, Han~\etal trained ChartLlama~\cite{han2023chartllama}, which demonstrates superior performance across various visualization tasks including NL2VIS, based on LLaVA~\cite{liu2024visual}.
Tian~\etal~\cite{tian2024chartgpt} broke down the process of visualization generation into step-by-step tasks and fine-tuned FLAN-t5 model~\cite{chung2022scaling} to align with the intended task. 
Although the above-mentioned methods demonstrate tremendous potential in generating various charts.
\revise{We} observe that their generated results still exhibit issues, ranging from code execution failures to incorrect data transformations, and missing legends.
Such shortcomings highlight the urgent need for systematic evaluation to understand the capability of LLMs and the performance of LLMs-based methods and gain insights for future advancements.



\subsection{Evaluation for Generated Visualization}
As LLMs demonstrate tremendous potential, researchers are increasingly engaged in assessing the quality of visualizations generated by LLMs~\cite{chen2023beyond,cheng2023gpt,kim2023good,pere2024are,podo2024vi}. 
A series of human evaluations were conducted from different aspects, including visualization code generation~\cite{cheng2023gpt,pere2024are}, visualization design~\cite{chen2023beyond,kim2023good}, and visual data exploration~\cite{chen2023beyond}.
Given the labor-intensive nature of human evaluation, automated evaluation methods~\cite{maddigan2023chat2vis1,li2024visualization,dibia2023lida, han2023chartllama,podo2024vi} show their importance in facilitating the iteration and improvement of methods, along with providing an objective assessment.
EvaLLM~\cite{podo2024vi} automates the evaluation of generated Vega-Lite visualizations in JSON format. It verifies the JSON structure, assesses code and JSON structure similarity, and compares data mapping, marks, and axes with the ground truth.
However, only charts written in high-level representation languages (e.g., Vega-Lite) are applicable to this method, and doubts remain about whether code similarity is a good measure.
Rule-based methods~\cite{maddigan2023chat2vis1,li2024visualization} focus on the presented chart and automatically check whether the data along the x-axis and y-axis matches the ground truth data.
However, they often overlook errors in channels other than the axes, such as using the same color to represent distinct categories, and they are sometimes too strict as the appropriate visual mapping can vary.
Other automated methods~\cite{dibia2023lida, han2023chartllama} employ self-evaluation strategies, wherein LLMs are utilized to assess the quality of code generated by themselves.
However, it remains unexplored whether LLMs possess the capability to determine the quality of generated visualizations solely based on code.

Most of the aforementioned methods only focus on ensuring that the visualization is ``correct'', yet a ``good'' chart should possess many other properties, such as readability~\cite{behrisch2018quality}, effectiveness~\cite{behrisch2018quality}, memorability~\cite{saket2016beyond}, and learnability~\cite{figueiras2018review}. 
Efforts within the visualization community have been made to automate the assessment of visualizations from these perspectives. 
For example, McNutt and Kindlmann~\cite{mcnutt2018linting} introduced a readability linter to assess visualizations created in Matplotlib against a set of rules.
To effectively scale up the coverage of evaluated aspects, recent machine learning-based approaches~\cite{fu2019visualization, wu2021learning} train models to asses the aesthetics, memorability, or layout quality of visualization images. 
However, all these methods solely look at the visualization itself, ignoring the original natural language query.
By introducing an end-to-end evaluation framework, we propose assessing the quality of generated visualizations alongside queries, thereby enhancing the comprehensiveness of the evaluation.



\subsection{Dataset for Visualization}
Datasets serve as a fundamental pillar in the benchmark.
However, previous NL2VIS datasets focused on narrow domains~\cite{gao2015datatone, kumar2016towards, srinivasan2021collecting}, which is not sufficient for comprehensive evaluation. 
For instance, NLV Corpus~\cite{srinivasan2021collecting} collected human annotator-written utterances for each visualization but was limited to three data tables.
\revise{Recently, several works have begun to collect large databases for visualization. However, they are primarily focused on real-world dataset gathering~\cite{viznet} or designed for specific visualization tasks such as visualization question answering~\cite{kim2020answering, masry2022chartqa, methani2020plotqa}, chart summarization~\cite{kantharaj2022chart, rahman2022chartsumm}, and visualization recommendation~\cite{hu2019vizml, dibia2019data2vis}. 
These datasets lack pairs of queries and visualizations, posing challenges for NL2VIS evaluation. }
\revise{The Quda dataset~\cite{fu2020quda}, which aims to advance the development of NL2VIS, comprises 14,035 user queries but lacks ground truth visualizations, further limiting its utility as a benchmark.
Another notable work, nvBench~\cite{luo2021synthesizing},} synthesized a dataset from an NL2SQL benchmark enabling NL2VIS model training and testing. 
This benchmark comprises 25,750 pairs of natural language queries and visualization, covering 105 domains \revise{across 153 datasets}.
Despite its extensive coverage, the dataset contains erroneous labels, along with the ambiguity of queries resulting in non-unique ground truth visualizations,  which could potentially lead to inaccuracies in evaluation results~\cite{maddigan2023chat2vis1,li2024visualization}.
We are still short of a large-scale dataset with accurately labeled ground truth to effectively benchmark LLMs. 
In this paper, we conducted a preliminary study to identify the key requirements for a benchmark dataset. 
\revise{Then we constructed a high-quality dataset with ground truth that includes chart types, plotted data, and meta-information to meet these requirements.}

%% file: sections/03-preliminaries.tex
\section{Preliminaries}

\subsection{NL2VIS Task Formulation}
In the era of large language models, a typical workflow of NL2VIS tasks involves assembling \textit{queries} along with \textit{tabular data} as input. 
Subsequently, LLMs are utilized to generate \textit{code} based on established visualization libraries, resulting in intermediary output. 
The generated code is executed in a sandboxed environment to obtain the final \textit{chart image}.
Hence, for the evaluation of visualizations produced by LLMs in NL2VIS tasks, it is imperative to assess the output of each stage, including the generated code and the resulting chart image.
We frame the evaluation scope within \revise{static charts, using widely adopted Python visualization libraries} such as Matplotlib~\cite{hunter2007matplotlib} and Seaborn~\cite{waskom2021seaborn}. 
These libraries are widely adopted within the community and are renowned for their versatility in producing a wide range of visualizations. 

\subsection{Preliminary Study}
\label{sec:pre-study}
To gain insights into the reliable evaluation of LLMs-generated visualizations, we conducted a preliminary study involving four human experts with more than five years of experience in the field of visualization.
Three hundred queries were randomly sampled from a widely used dataset nvBench~\cite{luo2021synthesizing}.
These queries were used to generate visualizations using four distinct LLMs: GPT-4, GPT-3.5, Gemini-Pro, and CodeLlama-7B.
Following this, four experts independently reviewed the codes and chart images generated by each of the four models, taking the ground truth provided in nvBench as a reference. 
Subsequently, a roundtable meeting was convened to discuss the findings of all evaluators. 
We made three observations throughout this process, regarding factors that are influencing the quality of generated visualizations and issues within the evaluation process. 

\textbf{Observation 1: Low-quality queries lead to nonsense results.} The selected three hundred queries have more deficiencies than anticipated.
This is primarily manifested when a query describes an unreasonable visualization, leading to visualizations that are either meaningless or confusing. For instance, identified data fields such as an ID were incorrectly treated as continuous numbers and mapped to the x-axis of a scatter plot.
Furthermore, some queries are ambiguous, making it difficult to ascertain if they meet the expected standards, and in some cases, the ground truth of certain queries is incorrectly labeled.
Upon discovering that the current generated results still exhibit numerous basic errors, we decided to focus our efforts on evaluating the generated visualization where queries explicitly define selected columns, aggregations, chart types, and order.
This clear type of query represents the most fundamental aspect of NL2VIS tasks.

\textbf{Observation 2: Inherent defects in the generated results.} We categorized the inherent defects that hinder comprehension into three dimensions based on expert discussions and previous literature related to visualization linting~\cite{chen2021vizlinter,mcnutt2018linting} and code generation benchmark~\cite{zhang2024benchmarking, lai2023ds}.

\noindent\underline{\textit{I1.~Invalid codes lead to rendering failure.}}
The generated code may be invalid for rendering visualizations, for example, crashing due to incorrect API usage or printing data instead of plotting. 

\begin{figure}[t]
    \centering    \includegraphics[width=\linewidth]{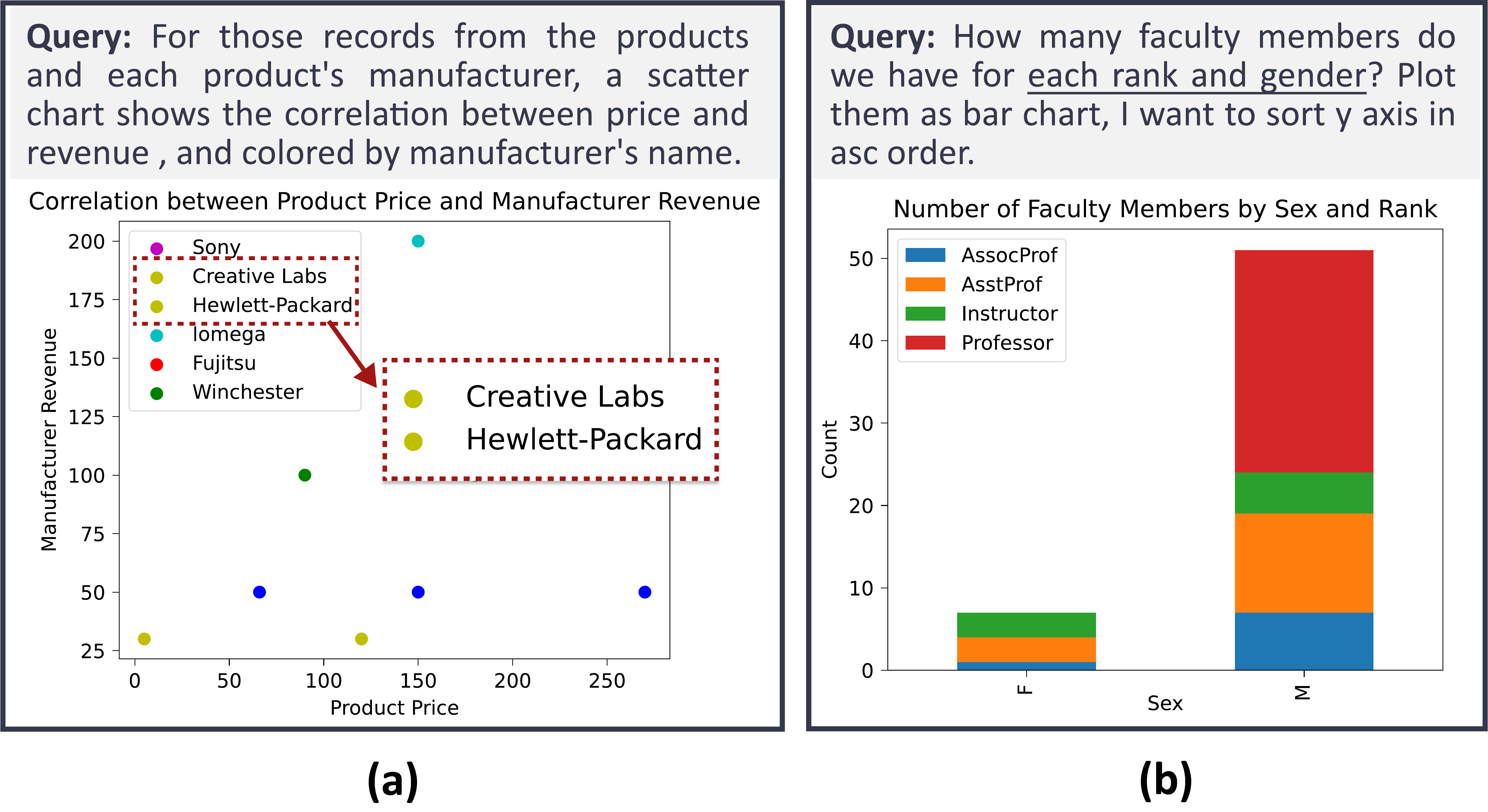}
    \vspace{-5mm}   
    \caption{Example cases where previous methods fail short: (a) the absence of consideration for color channels, leading to the oversight of identical colors \begin{tikzpicture}
  \fill[legend] (0,0) circle (0.1cm);
\end{tikzpicture} being used for different categories; and (b) misjudgment due to exact matching, where the ground truth maps the ``rank'' data field to the x-axis and the ``sex'' data field to the color channel. Since the query did not explicitly specify which data field should be mapped to which channel, this case should also be considered appropriate.}
    \label{fig:evalution_issues}
    \vspace{-5mm}
\end{figure}

\noindent\underline{\textit{I2.~Illegal charts do not meet requirements.}} The charts may be illegal due to conflicts with queries, such as selecting incorrect data columns and plotting inaccurate legends (e.g.,~\cref{fig:evalution_issues}(a)).

\noindent\underline{\textit{I3.~Low readability charts hinder comprehension.}} The charts face challenges that hinder comprehension due to various factors such as text overflow or overlap, low-contrast colors, and typographical errors.

\textbf{Observation 3: Low-effort and reliable evaluation is challenging.}
While the authors and the employed experts identify numerous issues through human evaluation, this methodology proves unsustainable for future benchmarking due to its labor-intensive nature and lack of objectivity.
Our preliminary attempts to reduce human effort and automate the evaluations are as follows.
Firstly, we adopted some rule-based automated methods~\cite{maddigan2023chat2vis1,li2024visualization}, and found them unsatisfactory. For example, they often compared data along the x and y axes directly with the ground truth, but sometimes the suitable visual mapping can be non-unique.
\cref{fig:evalution_issues} illustrates two cases where these methods fall short.
Secondly, we explored alternative methods~\cite{dibia2023lida, han2023chartllama} that utilize LLMs to score the code they generate, without considering the discrepancies between the resulting charts and the code. 
These preliminary efforts, utilizing prompts specified in prior studies and leveraging LLMs to identify issues, yielded limited success.

\subsection{Benchmark Requirements}
\label{sec:design_goals}
Therefore, we summarize three benchmark requirements as follows.

\noindent\textbf{R1. Incorporate a high-quality and large-scale dataset.} The dataset should demonstrate high quality, with accurately annotated ground truth and non-ambiguous, rational queries. Additionally, the benchmark dataset should be large in scale and cover a broad domain to ensure comprehensive evaluation results rather than one-sided assessments.

\noindent\textbf{R2. Support multi-dimension evaluation.} 
Given the potential for errors in both code and chart image, it is imperative that we systematically evaluate the visualizations for validity (\textit{I1}), legality (\textit{I2}), and readability (\textit{I3}). 
\revise{In this paper, we define validity as the ability of the code to render a visualization, legality as the compliance of the visualization with the query requirements, and readability as the effectiveness of the visualization in clearly presenting information.}
This multi-dimension evaluation ensures that we not only detect errors stemming from incorrect API usage or data mishandling but also identify issues related to readability.

\noindent\textbf{R3. Automate reliable \revise{evaluation}.} 
Automating evaluation not only facilitates rapid iteration and refinement of NL2VIS methods during the rapid development of large language models but also ensures objectivity in assessing quality.
Consequently, reliable evaluation \revise{results} emerge as crucial, playing a pivotal role in effectively guiding improvement directions and informing advancement.

%% file: sections/04-methodology.tex
\section{\name: A benchmark for data visualization}
Following the requirements outlined in Section~\ref{sec:design_goals}, 
we construct a high-quality and large-scale dataset (\textbf{R1}).
Building upon this dataset, we propose a novel NL2VIS evaluation framework that covers multiple dimensions (\textbf{R2}), including validity, legality, and readability, as illustrated in~\cref{fig:evaluator}.
To ensure the reliability (\textbf{R3}) of our automated methodology, we conducted quantitative experiments.

\subsection{Dataset Construction}
\label{sec:augmentation}
The NL2VIS benchmark dataset typically includes pairs of natural language queries (\textit{NL}) and corresponding visualizations (\textit{VIS}). The queries and their associated data tables serve as input for the NL2VIS task, while the visualizations represent the ground truth. 

Based on our preliminary study and related benchmark works~\cite{zhang2024benchmarking, lai2023ds,sui2023gpt4table}, we identified four requirements that the benchmark dataset should meet:
1)~\textit{Large-Scale coverage}: 
The dataset needs to include a substantial number of queries and databases from diverse domains to mitigate bias. Moreover, it is crucial to ensure balanced data distribution to prevent bias from specific databases.
2)~\textit{High-quality queries}: 
The queries in the dataset must be unambiguous and rational, explicitly specifying selected columns, aggregations, and chart types while describing rational visualizations.
3)~\textit{Accurate ground truth}: The ground truth data in the dataset should be accurately labeled and capable of precisely describing acceptable visualizations.
4)~\textit{Valuable query selection}: Exclude overly simplistic queries~\cite{sui2023gpt4table}, which are queries that the model can almost always answer correctly, as they offer limited value but increase evaluation cost.

Previous NL2VIS datasets either concentrate solely on narrow domains~\cite{gao2015datatone, kumar2016towards, srinivasan2021collecting} or lack ground truth~\cite{fu2020quda}, which differs from our requirement. nvBench~\cite{luo2021synthesizing} is the closest match to our needs, comprising 7,247 visualizations (\textit{VIS}) and 25,750 (\textit{NL}, \textit{VIS}) pairs from 153 databases. 
\revise{However, some of its queries are ambiguous, irrational, duplicated, and have incorrect ground truth. 
Therefore, we construct our dataset based on nvBench to meet the aforementioned requirements.}
The primary objective is to curate high-quality and unique queries, rectify ground truth inaccuracies, and augment meta-information while ensuring large coverage of databases.
The dataset construction process is delineated as follows (refer to Appendix~\ref{sec:dataset_construction_details} for more details). 




\textbf{High-quality queries selection.}
To select high-quality and non-duplicate queries from nvBench, we implement a rigorous selection procedure that integrates the insight of state-of-the-art LLMs and expertise from visualization experts. This procedure involves three distinct steps: rule-based, LLMs-based, and human-based selection.
Firstly, we devised and implemented eight rules for filtering and correcting queries. For instance, we filtered out visualizations (\textit{VIS}) that erroneously treated unique data such as IDs or codes as numerical values.
Secondly, we leveraged LLMs to alleviate the workload of human experts. 
Due to concerns about potential biases when relying solely on a single LLM, we chose three state-of-the-art LLMs (i.e., GPT-4, GPT-3.5, Gemini-Pro) to vote on whether the queries were ambiguous or irrational.
We adopted a majority rule strategy to select queries deemed high-quality by two or more LLMs.
Finally, human experts reviewed queries to ensure their clarity, rationality, and non-duplication. When experts encountered queries that did not meet the requirements, if the database and chart type associated with the query had multiple other instances, it was deleted directly. Otherwise, it was manually modified or rewritten to create a new query.

\begin{figure}[t]
    \centering    \includegraphics[width=\linewidth]{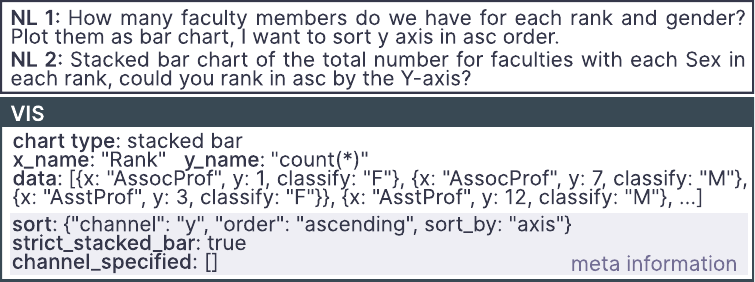}
    \vspace{-5mm}
    \caption{\revise{Example of (NL, VIS) pairs. Two NL queries correspond to the same VIS. Note that the ground truth VIS represents a feasible region for multiple acceptable visualization instances.}}
    \label{fig:dataset_example}
    \vspace{-1mm}
\end{figure}

\textbf{Accurate ground truth labeling.}
\revise{The ground truth in our dataset includes chart type, plotted data, and meta-information. This meta-information, which details implicit and explicit query specifics, serves as constraints during evaluation to determine the most appropriate charts. Human experts corrected the chart types and data, and added meta-information such as specified channels, sorting requirements, and whether grouped bar charts could be used instead of stacked bar charts. \cref{fig:dataset_example}
presents an example of the ground truth.}

\textbf{Dataset rebalancing.}
Motivated by previous research~\cite{sui2023gpt4table}, we implemented a filter to exclude \revise{overly simple queries, which are predictable and universally solvable by most models, thereby limiting their evaluative utility.}
Simple queries are those for which GPT-4, GPT-3.5, Gemini-Pro, and CodeLlama-7B can generate correct answers in a zero-shot setting (as elaborated in Section~\ref{sec:setup}).

\begin{figure}[t]
    \centering    \includegraphics[width=\linewidth]{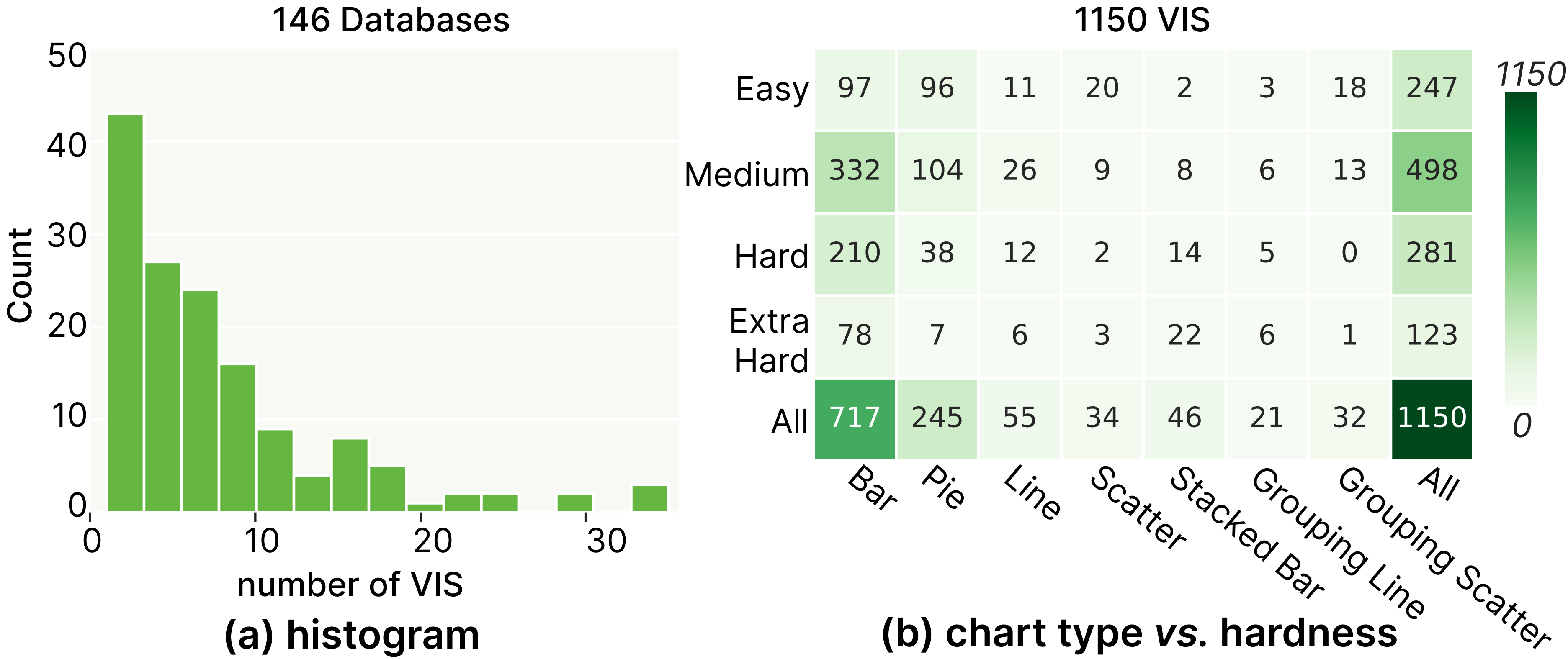}
    \vspace{-6mm}
    \caption{Statistical analysis of the dataset: \revise{(a) A histogram of the number of visualizations per database}, and (b) the distribution of visualizations across different chart types and hardness.}
    \label{fig:benchmark_statistic}
    \vspace{-6mm}
\end{figure}

We end up with 1,150 distinct visualizations (\textit{VIS}) and 2,524 (\textit{NL}, \textit{VIS}) pairs, covering 146 databases. 
 Considering the inherent flexibility of language, we preserved multiple (\char`\~2.19) NL queries describing the same VIS, treating them as a cohesive entity during evaluation.
We adhered to the hardness definition established in nvBench, which pertains to the complexity associated with chart generation. More precisely, visualizations are classified into four distinct levels of hardness: easy, medium, hard, and extra hard.
\cref{fig:benchmark_statistic} presents the statistics of our dataset, providing insights into coverage and diversity. 

\subsection{Evaluation Framework Overview}
\label{sec:framework}

\begin{figure*}[t!]
    \centering    \includegraphics[width=0.85\linewidth]{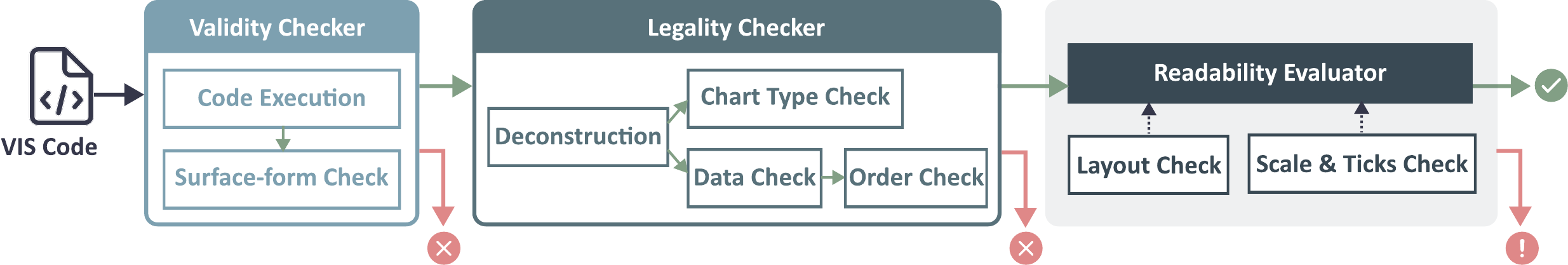}
    \vspace{-3mm}
    \caption{The pipeline of \name includes three key modules: the validity checker, the legality checker, and the readability evaluator. }
    \label{fig:evaluator}
    \vspace{-5mm}
\end{figure*}

To address the requirements outlined in \textbf{R2} and \textbf{R3}, we design and illustrate the pipeline of our evaluation framework in~\cref{fig:evaluator}.
The code generated by LLMs undergoes automated assessment for validity, legality, and readability to identify any potential issues within the visualization.
We describe the construction of each checker as follows.


\subsubsection{Validity Checker}
The validity checker verifies whether the code can render visualizations through two steps. First, it executes the code in a sandboxed environment to ensure that no crashes occur during execution. Then, it performs a surface-form check to verify if the code contains the necessary code snippets to render visualizations (e.g., \texttt{plt.show()}).

\subsubsection{Legality Checker}
The legality checker verifies whether the generated visualization meets the query. 
It begins by deconstructing the visualization to extract information (e.g., chart type, data).
Then a series of impartial checks examine the chart type, data, and order with the aid of meta-information.

After successfully rendering a visualization, the resulting image is saved in Scalable Vector Graphics (SVG) format. SVG, as a vector format, is well-suited for parsing the information contained within the visualization. 
While some methods exist for extracting data or analyzing chart types from raster images~\cite{poco2017reverse, zhou2023enhanced, han2023chartllama}, they are not robust.
Therefore, we adopt SVG-based deconstruction~\cite{wu2020mobilevisfixer, chen2023mystique}.
The logical structure inherent in images created by visualization libraries like Matplotlib allows us to precisely extract plotted data and parse additional information such as chart type, axes, and legends through the ``id'' attribute. Here, ``id'' refers to the unique identifiers utilized in SVG.
Specifically, our deconstruction process cannot handle unconventional charts, such as those with dual axes or irrational data mappings that result in missing ticks (e.g., ~\cref{fig:casestudy}~(7)). 
\revise{Due to our practical findings indicating that these charts are illegal, we categorize them as such by default. However, they are labeled as ``unparseable'', allowing for human verification as needed for thoroughness.}   

With accurately labeled ground truth and meta-information, we can reliably evaluate whether the chart type, presented data, and the order in the generated visualization meet the requirements. 
We evaluate the sorting order separately from the data for two main reasons. Firstly, identical values in the data can lead to multiple valid sorting outcomes, resulting in potential errors if data is directly compared. Secondly, visual sorting refers to actions applied to the resulting charts, not the underlying data itself~\cite{tian2024chartgpt}. For instance, in a stacked bar chart, the order is determined not directly by the y-axis values but by the sum of the stacked bars. \revise{Details about how our method verifies whether the generated visualization meets the query are provided in Appendix~\ref{sec:legality_checker}.}

\subsubsection{Readability Evaluator}
\label{sec:readability_checker}

To evaluate the readability of generated visualizations \revise{in the context of their queries}, we propose an innovative assessment methodology integrating a multimodal model named GPT-4V(ISION)\cite{openai2023gpt4v}. 
GPT-4V exhibits remarkable capabilities in processing both textual and image inputs and generating textual outputs\cite{shi2023exploring,abe2024assessing}.
We conducted a pilot experiment to further understand the capabilities of GPT-4V. We presented GPT-4V with the visualization and posed multiple questions such as ``Whether any graphic elements out of the canvas?'' and ``Any readability issues within this visualization?''. 
The model showed impressive proficiency in text recognition and comprehension of visualizations.
However, it occasionally made simple errors, such as failing to identify overflow issues where text or legends extend beyond the canvas or detecting only partial problems.

Therefore, we decided to decompose the complex readability assessment problem into smaller and more controllable sub-problems, which is a common strategy to mitigate errors in LLMs~\cite{zhou2022least}. We identified two significant readability-related issues and prioritized them as sub-problems to address. As depicted in~\cref{fig:readability}, the readability assessment will be conducted with the help of layout check and scale \& ticks check. The details of each module are as follows:

\begin{figure}[t]
    \centering    \includegraphics[width=\linewidth]{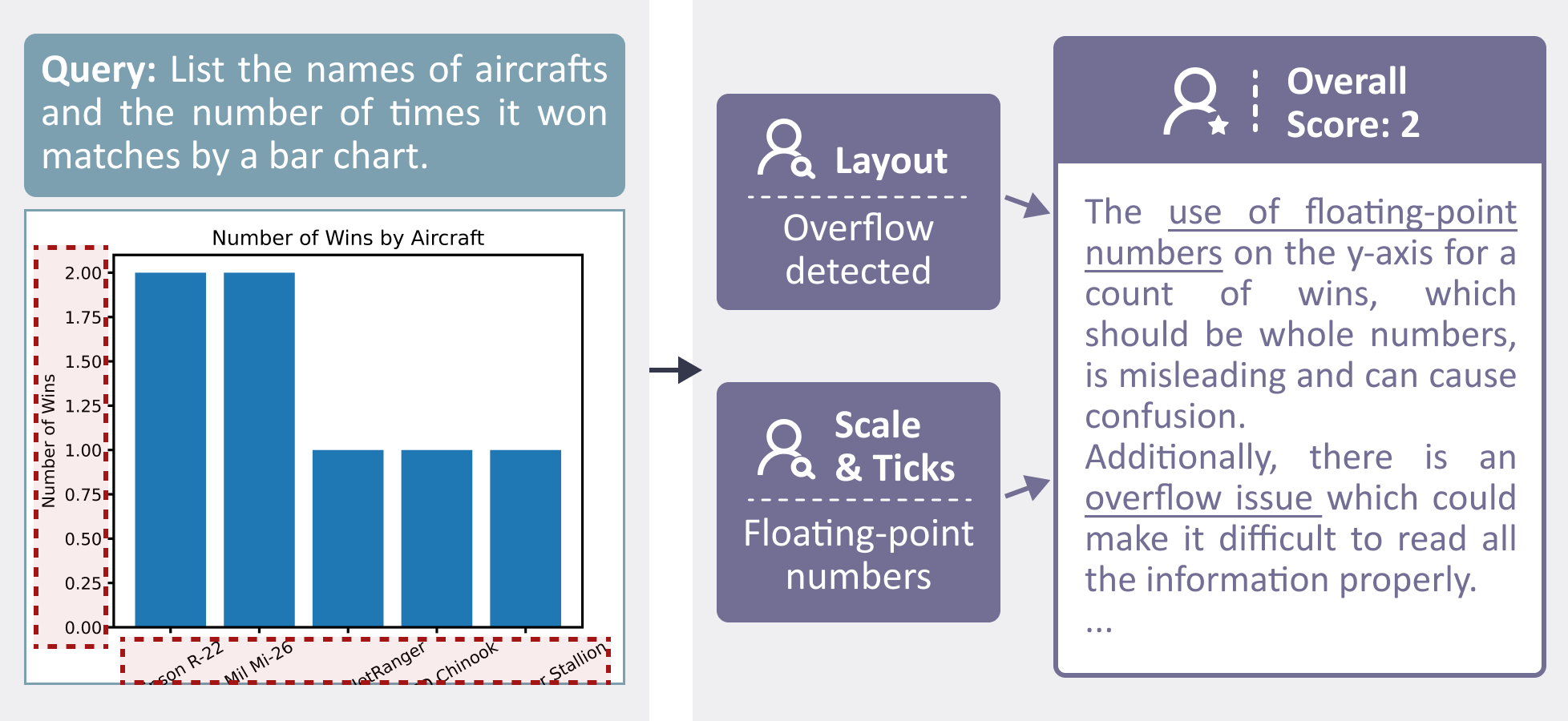}
    \vspace{-6mm}
    \caption{An example of using the readability evaluator. The layout check identified issues with the overflow of ticks and the title on the x-axis. The scale \& ticks check revealed that the y-axis ticks were displayed using floating-point numbers, which is unconventional for representing integer values like the count of wins. These evaluations were given to the readability evaluator, resulting in a final overall score of 2 along with a concise rationale.}
    \label{fig:readability}
    \vspace{-3mm}
\end{figure}

\textbf{Layout Check.}
The layout check entails assessing overflow and text overlap within visualizations. We conducted experiments using various prompt strategies and observed that GPT-4V's accuracy in this task was not sufficiently high to be incorporated into the evaluation framework.
At times, words partially extended beyond the canvas, with segments of the letters fully displayed, posing a challenge to the model.
Moreover, there were instances where the model could infer complete words from partial text, potentially influencing its ability to accurately judge whether words were fully displayed. 
Consequently, we opted for a more reliable approach by simulating a browser environment. This methodology allows us to precisely determine the size of the canvas and the size and position of visualization elements in SVG format, facilitating an accurate assessment of overflow and overlap.

\textbf{Scale \& Ticks Check.}
The scale \& ticks check aims to determine if the chosen scale is suitable for interpreting values, avoiding unconventional scales such as an inverted y-axis scale. Additionally, it assesses the appropriateness of the displayed ticks when evaluating axes, avoiding unconventional choices such as representing years with floating-point numbers.
One notable observation is the phenomenon of hallucination that occurs when GPT-4V interprets ticks; integer values may be inaccurately perceived as floating-point values.
To address this, we incorporate deconstructed ticks from the x-axis and y-axis as auxiliary information in the prompt provided to GPT-4V. 
This inclusion aids the model in conducting more precise evaluations and reducing the potential for hallucination.

\textbf{Overall Readability Rating.}
The overall readability rating systematically evaluates the readability of visualizations, considering various factors beyond layout and scale, such as title, labels, colors, etc., assigning scores from 1 to 5 points.
A score of 1 denotes that the visualization is highly challenging to comprehend, while 5 indicates that it is very easy to comprehend.
As previously described, both the layout check and the scale \& ticks check provide assessments on potential issues within their respective domains, accompanied by concise justifications for their evaluations. These evaluations are then integrated into the prompt for the overall readability rating.
The prompt also includes the query, enabling more precise judgments by aligning with the specific demands of the visualization. For instance, it facilitates verification of the relevance and clarity of information presented in the title and labels.
An important observation is that the model often exhibits skepticism regarding the accuracy of data and sorting, frequently perceiving visualizations with correct sorting as not meeting the specified requirements. Thus, we emphasize in the prompt, ``Do not consider the correctness of the data and order in the visualizations, as they have already been verified.'' For the detailed prompt, please refer to Appendix~\ref{sec:readability_evaluator}.

\subsubsection{Implementation}
Following the framework described above, we develop a Python package, \texttt{\name}\footnote{Source code and dataset are available at \url{https://github.com/microsoft/VisEval}.}, which embeds a function \texttt{evaluate()} to evaluate generated visualization with all available checkers. 
\revise{Our evaluation framework is designed with modularity, making it easy to configure according to user requirements. In cases where a vision model isn't configured, the framework checks aspects that don't depend on a vision model. Users also have the flexibility to} independently check specific aspects according to their interests; for instance, using the function \texttt{readability\_evaluate()} will solely evaluate readability.

\subsection{Quality Assurance}
To ensure the quality of our benchmark, each query, ground truth, and module within the evaluation framework underwent scrutiny by experts experienced in data visualization. 
We meticulously designed test cases to thoroughly assess the validity and legality checker. 
Furthermore, we evaluated the performance of our readability evaluator by collecting ratings from three human experts and quantitatively measuring the quality as follows.

\underline{\textit{Data preparation.}} 
One hundred visualizations generated by GPT-4, GPT-3.5, Gemini-Pro, and CodeLlama-7B were randomly sampled. 
Detailed information about the generation is provided in Section~\ref{sec:setup}.
Three experts, each having more than five years of experience in the field of visualization, independently rated the readability of each visualization using a scale ranging from 1 to 5. Subsequently, the average score for each visualization was calculated based on their independent ratings.
Furthermore, the experts convened to collectively analyze whether the visualizations presented specific layout issues such as overflow or overlap, in addition to scale and ticks issues. These identified issues were then marked using boolean values to indicate their presence.

\underline{\textit{\revise{Metrics.}}}
We naturally treat layout check and scale \& ticks check as a classification problem, evaluating them based on accuracy rate.
We quantify the consistency between automated ratings and human ratings using Spearman's rank correlation coefficient (SRCC), a widely acknowledged metric utilized in previous assessment studies~\cite{fu2019visualization, borkin2015beyond}. 
\revise{SRCC measures how well the order of predicted ratings matches the order of ground truth ratings, within the [-1, +1] range where 1 indicates the predicted ratings are perfectly ordered the same as the ground truth and -1 means they are ordered in exactly the opposite way.}

\underline{\textit{Result analysis.}} 
\revise{In summary, the layout check attains a perfect accuracy rate of 100\%. The scale and ticks check achieves a 99\% accuracy rate. However, without including deconstructed ticks as auxiliary information, the accuracy decreases to 92\%.}
Our readability rating method obtains an impressive SRCC score of 0.843, indicating a significant correlation with human experts.
Additionally, we observed an average SRCC of 0.782 among pairs of experts, further highlighting the reliability of our method for readability assessment and comparative analysis. 
\revise{The correlation between readability evaluator ratings and average human ratings is illustrated in~\cref{fig:readability_correlation}. After analysis of the evaluators' rationales and expert interviews, variability was observed in perceptions regarding the impact of different visualization issues on readability. Further discussions will be detailed in the Appendix~\ref{sec:readability_discussion}.}

Furthermore, we conducted an ablation study to assess the influence of including the layout check or scale \& ticks check in the readability evaluator, as detailed in ~\autoref{tab:compare_checker}.
The analysis results indicate that excluding checks leads to a decrease in token count, but also a reduction in correlation. 
In order to ensure reliable evaluation outcomes, we retain both checks in our readability evaluator.


\begin{table}[ht]
    \centering
    \begin{minipage}{0.50\linewidth}
        \centering
        \includegraphics[width=0.96\linewidth]{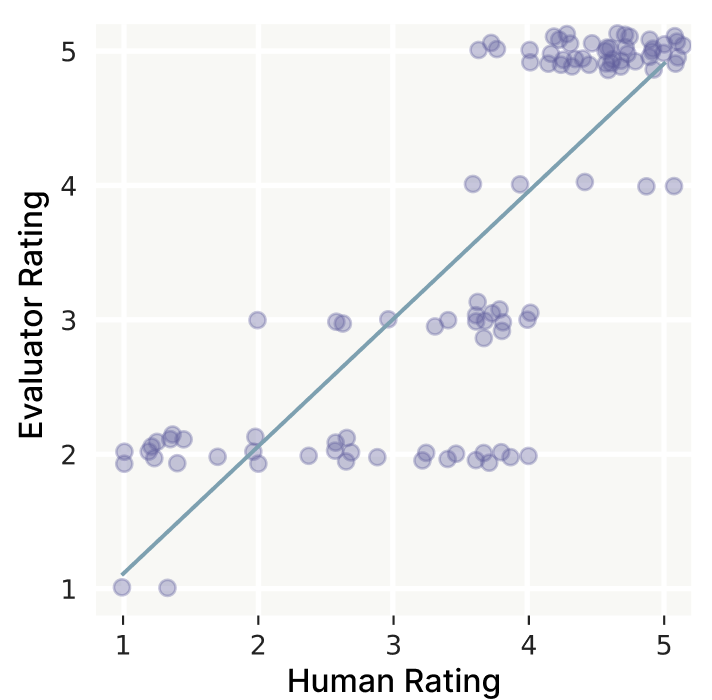} 
        \vspace{-2mm}
    \captionof{figure}{\revise{Correlation between human and readability evaluator ratings. Points are slightly jittered to prevent overlap.}}
    \label{fig:readability_correlation}
    \vspace{-3mm}
    \end{minipage}
    \hspace{0.02\linewidth} 
    \begin{minipage}{0.45\linewidth}
        \caption{Comparison of SRCC and consumed prompt tokens among different readability evaluator prompts. We compare: default, without scale \& ticks check, without layout check, and without both checks.}
        \vspace{-2mm}
        \resizebox{\linewidth}{!}{%
            \begin{tabular}{ccc}
            \toprule
             Prompt & SRCC & \# Tokens \\
            \midrule
            default & \bfseries 0.843  & 2073.38\\
            w/o scale \& ticks & 0.732& 1071.02 \\
            w/o layout & 0.675 & 2063.72 \\
            w/o both & 0.507 & 1057.86\\
            \bottomrule
            \end{tabular}
        }
        \label{tab:compare_checker}
    \end{minipage}%
\end{table}

%% file: sections/05-evaluation.tex
\section{Evaluation}
\label{sec:evaluation}

\subsection{Setup}
\label{sec:setup}
\textbf{NL2VIS prompt.}
The prompt significantly influences the performance of pre-trained language models~\cite{zhao2021calibrate}, making its selection crucial. 
In real-world scenarios, data is not always well-organized, leading to 461 out of 1150 visualizations in our dataset being generated based on multiple tables.
However, LLMs-based visualization generation methods, to our knowledge, do not support generation from multiple tables.
CoML\footnote{\url{https://github.com/microsoft/CoML}}~\cite{zhang2023mlcopilot}, a data science code generation method, caught our attention for its remarkable capabilities~\cite{zhang2024benchmarking} and capability to generate code from multiple tables.
We decided to revise its few-shot prompt~\cite{kaplan2020scaling} to specifically focus on visualization generation.

\begin{figure}[t!]
    \centering    \includegraphics[width=\linewidth]{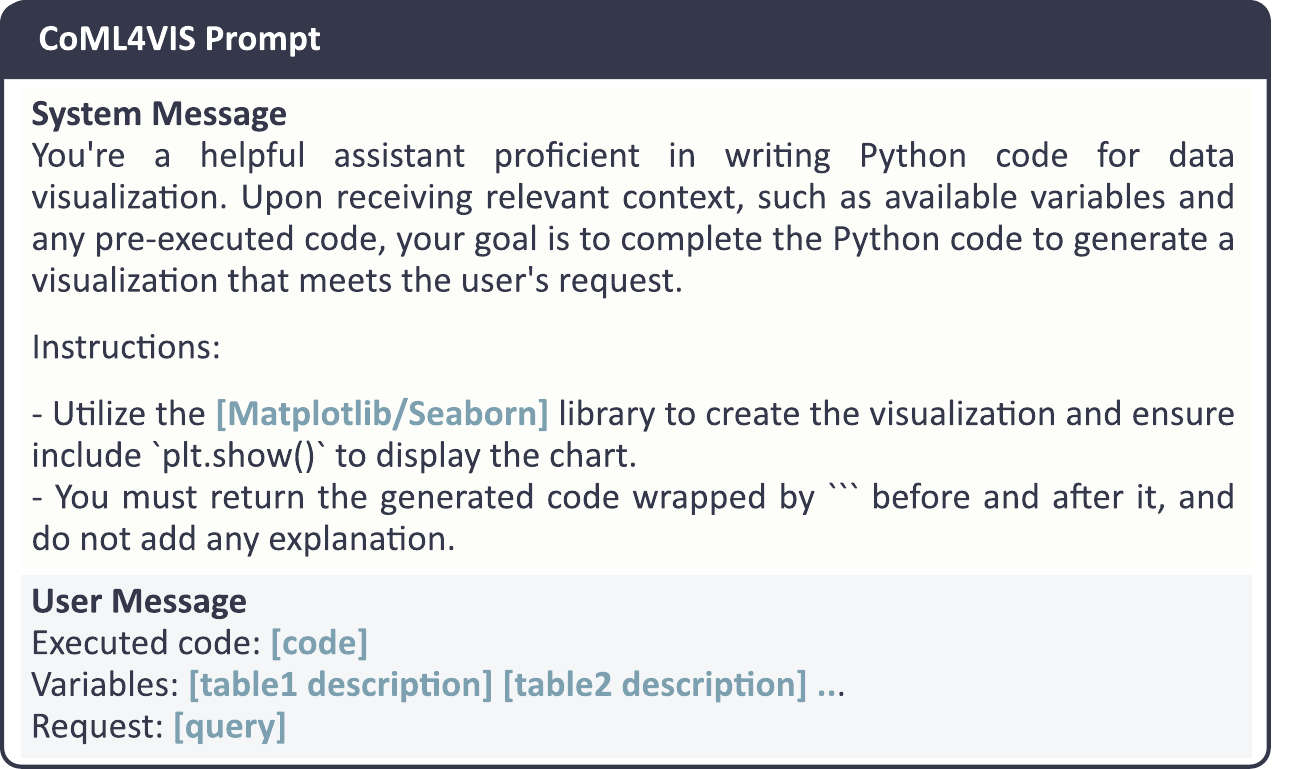}
    \vspace{-6mm}
    \caption{The prompt template for CoML4VIS. }
    \label{fig:coml}
    \vspace{-5mm}
\end{figure}

As depicted in ~\cref{fig:coml}, the prompt begins with a task description and instructions, followed by executed code, table descriptions, and a natural language query. The executed code includes package imports and operations for reading tables. By default, only the tables required for generating visualizations are accessed. 
The table descriptions provide a summary of the information contained in the accessed tables, detailing column names and samples of $N$ rows (where $N=10$ in our evaluation), as shown in~\cref{fig:table_formatter}(a).
In line with prior research~\cite{zhang2023mlcopilot}, which highlighted significant improvements with a single example compared to a zero-shot setting, we chose a bar chart example that integrates data from two tables as our one-shot example. 
To enhance the chart's readability, we rotated the ticks and adjusted the ticks to display integers. More detailed information about \revise{the prompt design} is provided in \revise{Appendix~\ref{sec:evaluation_choices}}.
We refer to this revised visualization generation approach as \textbf{\textit{CoML4VIS}} in the following sections.

\noindent\textbf{Visualization library.} We compare two well-known Python visualization libraries: Matplotlib~\cite{hunter2007matplotlib} and Seaborn~\cite{waskom2021seaborn}. They differ in their level of abstraction and ease of use for creating various types of plots.
Matplotlib provides an extensive range of plotting options and customization capabilities. Seaborn is built on top of Matplotlib and offers a higher-level interface for creating visualizations with fewer lines of code, thus making it easier to generate aesthetically pleasing and informative plots.
We specify the names of the selected libraries in the prompt, as shown in~\cref{fig:coml}.

\noindent\textbf{Models.} We evaluate the performance of four state-of-the-art models: GPT-4~\cite{openai2023gpt4}, GPT-3.5~\cite{ouyang2022training}, Gemini-Pro~\cite{team2023gemini} and CodeLlama-7B~\cite{roziere2023code}. Specifically, We set the hyper-parameter temperature to 0.

\noindent\textbf{Metrics.}
The ``\textit{Quality Score}'' provides a holistic assessment of the quality of generated results. 
\revise{If a result is determined to be invalid or illegal, its quality score is assigned a value of 0. Otherwise, the quality score is equal to the readability score assigned by the readability evaluator.}
Since multiple queries may correspond to the same visualization instance, the overall quality score for each visualization instance is calculated as the sum of individual scores divided by the total number of queries for that instance.
Furthermore, we ascertain the ``\textit{Pass Rate}'' as the ratio of valid or legal results to the total number of queries, excluding the readability score from this calculation to accommodate less stringent scenarios.
For a more nuanced analysis, error rates can be separately computed for validity and legality, denoted as the ``\textit{Invalid Rate}'' and ``\textit{Illegal Rate}'', respectively. 
Additionally, the ``\textit{Readability Score}'' \revise{is calculated as the average readability score for visualizations that have been assessed for readability, which are only those that are valid and legal.}

\subsection{Main Results}
\noindent\underline{\textit{Quality score:}} 
\autoref{tab:compare_models} displays the quality score across four LLMs. We find that \name can differentiate models with different capabilities. 
The top-performing model GPT-4 achieves a quality score of 2.89 in the Matplotlib setting and 2.31 in the Seaborn setting, with the optimal score being 5. These scores are non-trivial but fall short of perfection, indicating that there is room for improvement.
Other models have lower quality scores, and the ranking of the models is approximately as follows: CodeLlama-7B < Gemini-Pro < GPT-3.5 < GPT-4.
Contrary to our expectations, when using Seaborn, all models do not achieve a higher quality score. 
We observed an increase in their invalid rate, particularly GPT-4, which experienced the largest increase (22.12\%). 
This indicates that their pre-training corpus may have less content related to Seaborn compared to Matplotlib, leading to greater challenges in generating code that can render visualizations.

\input{tables/comparision_llms}

\begin{figure}[t!]
    \centering    \includegraphics[width=\linewidth]{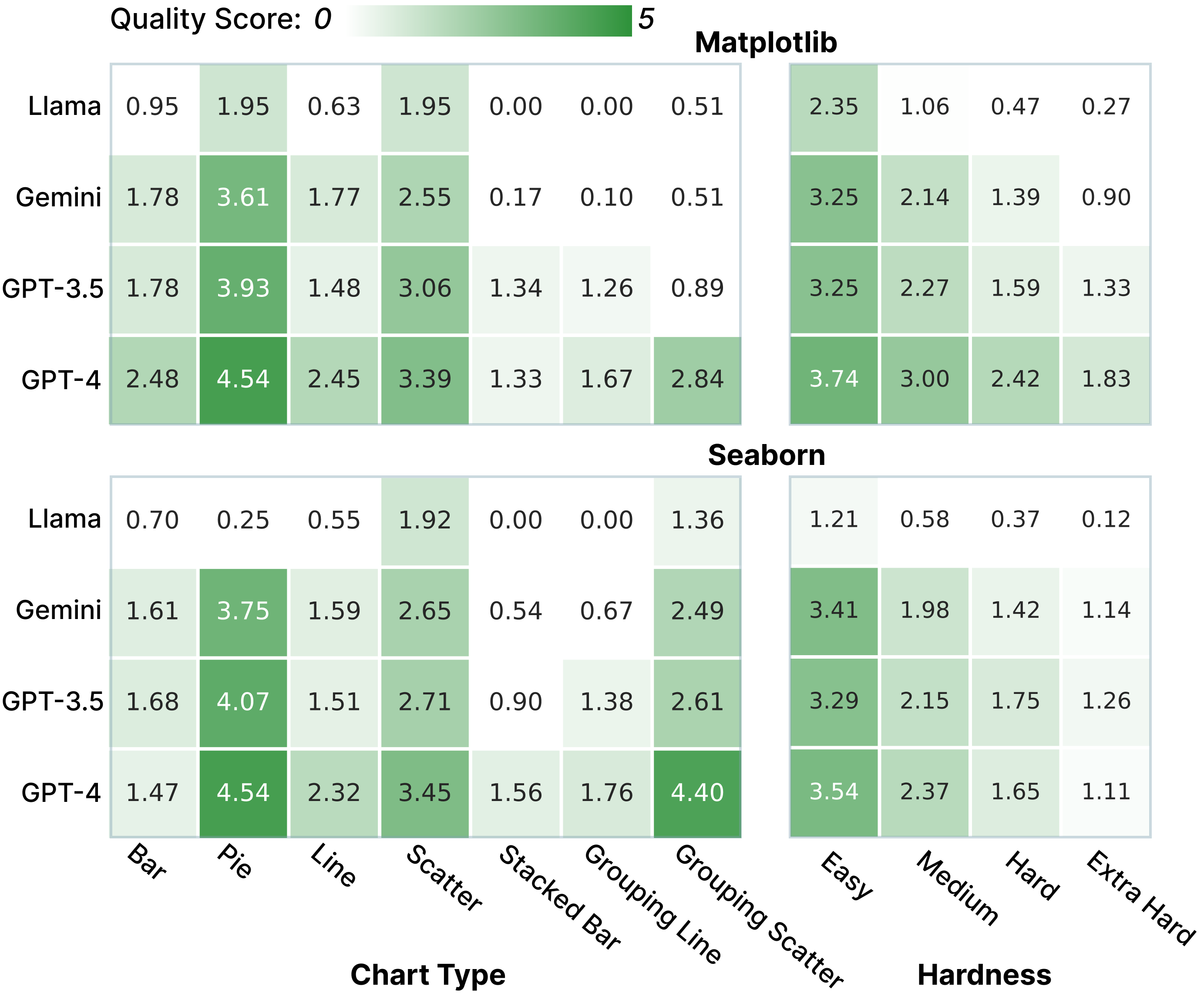}
    \vspace{-5mm}
    \caption{Quality score on different chart types and hardness across LLMs. Llama refers to CodeLlama-7B and Gemini refers to Gemini-Pro.}
    \label{fig:chart_hardness_result}
    \vspace{-3mm}
\end{figure}

\noindent\underline{\textit{Different chart type:}} 
LLMs exhibit varying performance across different chart types, as illustrated in \cref{fig:chart_hardness_result}. 
Charts requiring three visual channels (i.e., stacked bar charts, grouping line charts, and grouping scatter plots) tend to have lower quality scores compared to charts of the same type that only require two visual channels (i.e., bar charts, line charts, and scatter plots). 
This observation suggests that LLMs encounter difficulties when handling complex visualizations.

\noindent\underline{\textit{Different hardness:}} 
The complexity of chart generation, as reflected in hardness, also impacts the quality of the generated chart, as depicted in~\cref{fig:chart_hardness_result}. This trend is observed across all four models, with the quality score decreasing as the complexity of chart generation increases.

\begin{figure}[t!]
    \centering    \includegraphics[width=\linewidth]{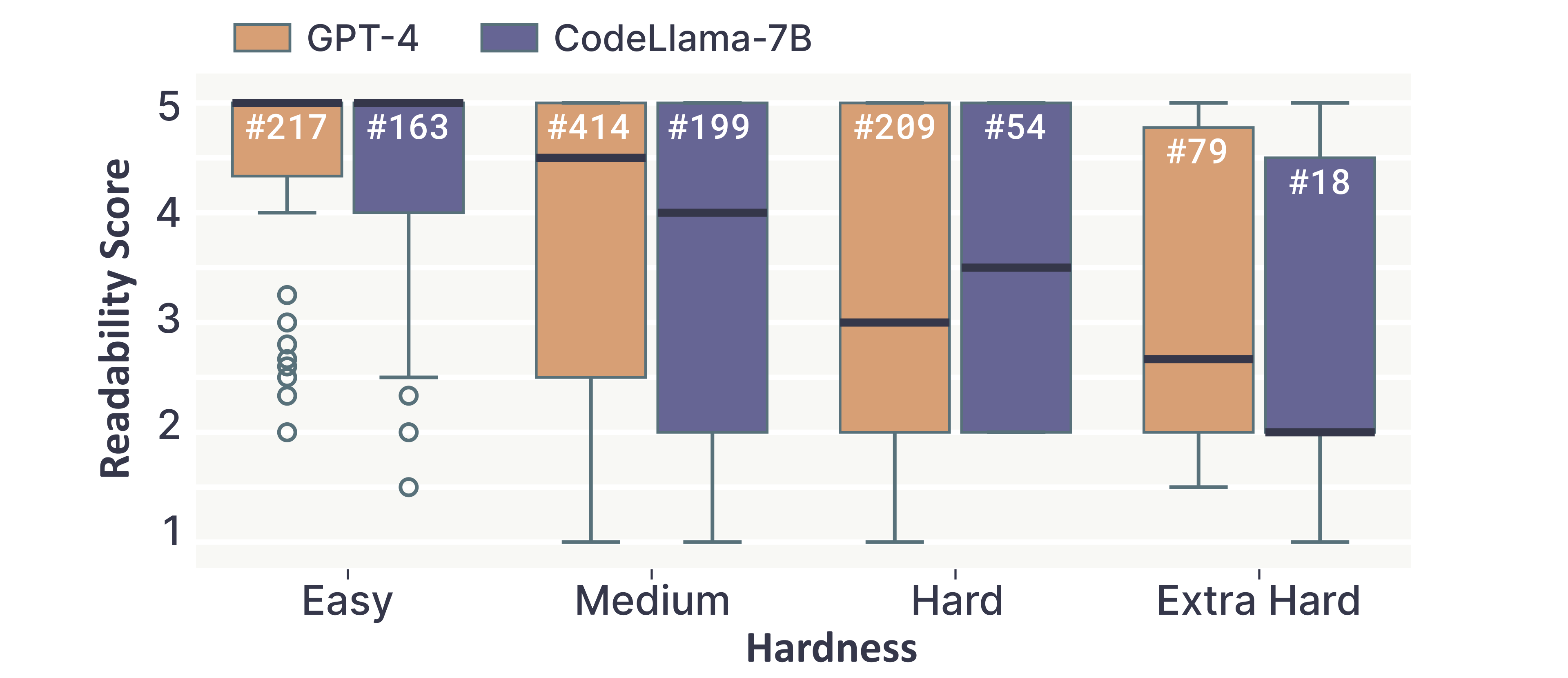}
    \vspace{-5mm}
    \caption{Comparison of the readability scores between GPT-4 and CodeLlama-7B across different hardness using Matplotlib library. The value represented by ``\#'' in the boxplot indicates the number of visualizations evaluated for readability scores.}
    \label{fig:readability_score}
    \vspace{-1mm}
\end{figure}

\noindent\underline{\textit{Readability score:}} 
Despite achieving a strong readability score of 3.87, CodeLlama-7B exhibited the lowest pass rate at 28.17\% in the Matplotlib setting, as indicated in~\autoref{tab:compare_models}.
To investigate this discrepancy, we compared its readability score with that of GPT-4, which had the highest pass rate in our evaluation. 
\revise{As depicted in~\cref{fig:readability_score}, there is a trend of decreasing readability scores with increasing query hardness. Only visualizations passing the validity and legality checks are evaluated for readability. Given that GPT-4 has a larger assessment size than CodeLlama, its overall readability score tends to be lower.
However, when we focus on a subset of 415 visualizations that passed both validity and legality checks from GPT-4 and CodeLlama-7B, the readability score of GPT-4 (4.04) was higher than that of CodeLlama-7B (3.92).}
This underscores the significance of prioritizing the quality score for a comprehensive assessment of the overall performance of generation methods, even though the readability score is a useful indicator of the generated visualization's quality in terms of readability.


%

\begin{figure*}[t!]
    \centering    \includegraphics[width=\linewidth]{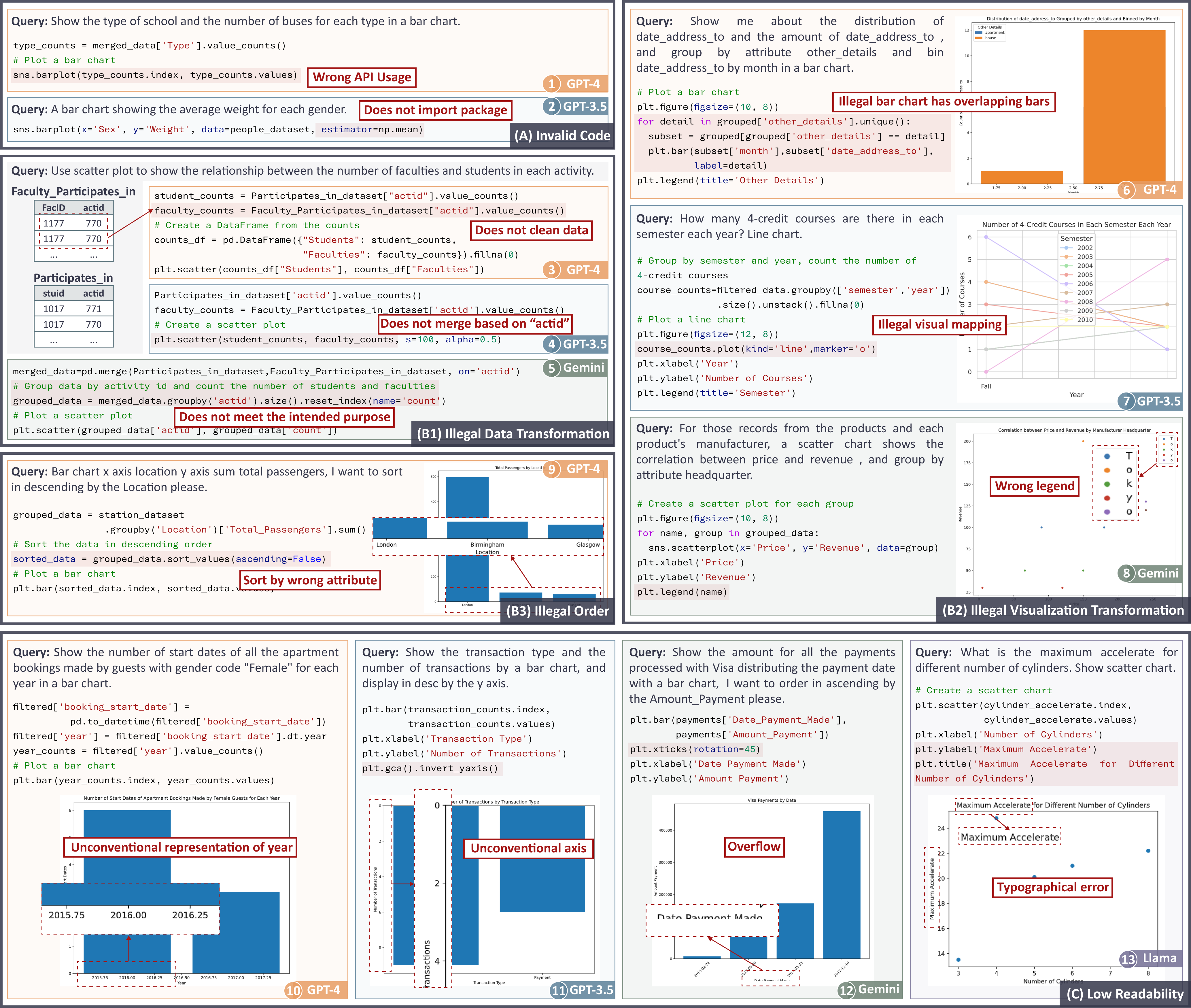}
    \vspace{-5mm}
    \caption{Typical errors: (A) pertains to invalid code error. (B1-3) denote illegal errors occurring during data transformation, visualization transformation, and sorting processes. (C) relates to issues of low readability.}
    \label{fig:casestudy}
    \vspace{-5mm}
\end{figure*}

\subsubsection{Typical Errors}
To gain a better understanding of the errors encountered when generating visualizations using LLMs, we \revise{manually} analyze the errors identified by each sub-check module (see Appendix~\ref{sec:typical_errors} for more details).
We then categorize these typical errors into five categories to reveal prevalent challenges.
\cref{fig:casestudy} demonstrates examples of errors from each category, elaborated upon in detail below.

\textbf{Invalid code.} 
Here are common reasons for such errors to occur: incorrect API calls or calling non-existent APIs, as illustrated in~\cref{fig:casestudy}(1) where \texttt{barplot()} is given two positional arguments despite accepting zero to one argument; forgetting to import packages, as depicted in~\cref{fig:casestudy}(2) where ``numpy'' (\texttt{np}) is used without being imported; hallucinations leading to the use of non-existent data columns.

\textbf{Illegal data transformation.} This type of error refers to instances where the chart does not meet the query due to incorrect data transformations. 
The primary reason for such errors is incomprehension of the data table. For instance, in~\cref{fig:casestudy}(3), two duplicate rows in the \texttt{Faculty\_Participates\_in} table should be cleaned to avoid double-counting faculty members with the same \texttt{FacID}, which GPT-4 does not handle.
Another reason is generating code that does not meet its intended purpose. For instance, in~\cref{fig:casestudy}(5), the comment mentions counting the number of students and faculties, while the code calculates the record count within each \texttt{actid} group.

\textbf{Illegal visualization transformation.} The third category indicates that the generated visualization does not undergo appropriate visual transformation. 
Some instances fail to create the legal chart types. 
This not only includes instances of mismatched chart types but also covers situations like the one shown in~\cref{fig:casestudy}(6), where bars were overlapped.
Additionally, there are instances involving improper visual mapping. 
In severe cases, this can lead to uninterpretable charts such as ~\cref{fig:casestudy}(7). 
Moreover, some instances involve forgetting to add a legend or creating an incorrect legend. 
For example, as depicted in~\cref{fig:casestudy}(8), where the ``name'' variable was erroneously included in the legend. 

\textbf{Illegal order.} Sorting issues resulting from a lack of sorting or incorrect sorting criteria. For instance, in~\cref{fig:casestudy}(9), the query specifies descending order based on the ``Location''. However, the generated code sorts based on 
\texttt{values} (the sum of total passengers) instead.

\textbf{Low readability.} Readability issues are common in the generated visualizations. For example, in~\cref{fig:casestudy}(10), representing years using decimals may confuse readers. In~\cref{fig:casestudy}(11), inverting the y-axis places the origin at the top-left corner, which is not consistent with common reading habits. Additionally, in~\cref{fig:casestudy}(12), the x-axis title overflows, leading to text truncation. Lastly, in~\cref{fig:casestudy}(13), there is a typographical error; it should be ``Acceleration'' instead of ``Accelerate''. These issues, to varying degrees, affect the understanding of the charts.

\subsection{Evaluating Other Approaches}
In this subsection, We demonstrate the performance of previous LLMs-based approaches, including Chat2vis~\cite{maddigan2023chat2vis} and LIDA~\cite{dibia2023lida}, as well as CoML4VIS.
Since both Chat2vis and LIDA are limited to generating visualizations from a single table, our comparison focuses on the quality of 689 visualizations that require data from a single table.
As summarized in~\autoref{tab:compare_approach}, CoML4VIS has the highest quality score, while Chat2vis has the highest pass rate, which shows that different prompt strategies can have varying effects on the model's performance.

\input{tables/comparision_approach}

\begin{figure}[t!]
    \centering    \includegraphics[width=\linewidth]{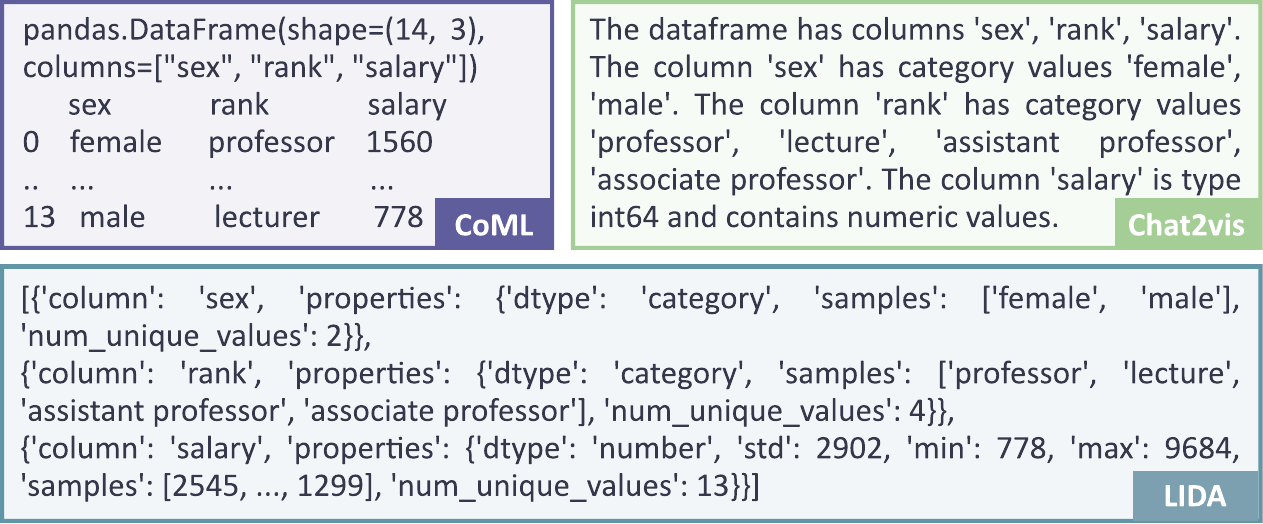}
    \vspace{-6mm}
    \caption{Illustration of table format in CoML, LIDA, and Chat2vis.}
    \label{fig:table_formatter}
    \vspace{-2mm}
\end{figure}

\revise{Additionally, we noticed that previous approaches introduced distinct table formats, so we conducted an evaluation to understand how the pass rate of the same model varies across these different formats.}
As depicted in ~\cref{fig:table_formatter}, CoML summarizes the column names and samples $N$ rows of data. LIDA describes the statistical information of each column in JSON format and samples $N$ values randomly for each column. Chat2vis uses natural language to describe the type of each column and provides $N$ examples for categorical data.
To ensure fairness, we maintained all other settings of CoML4VIS unchanged except for the table format, and we standardized the sample size $N$ to 10 for all formatting options.
We found that when using the table format of Chat2vis in CoML4VIS, the pass rate for visualizations requiring a single table can reach 70.43\%, which is 0.5\% higher than Chat2vis' pass rate and 2.88\% higher than the original CoML4VIS.

\begin{figure}[t!]
    \centering    \includegraphics[width=\linewidth]{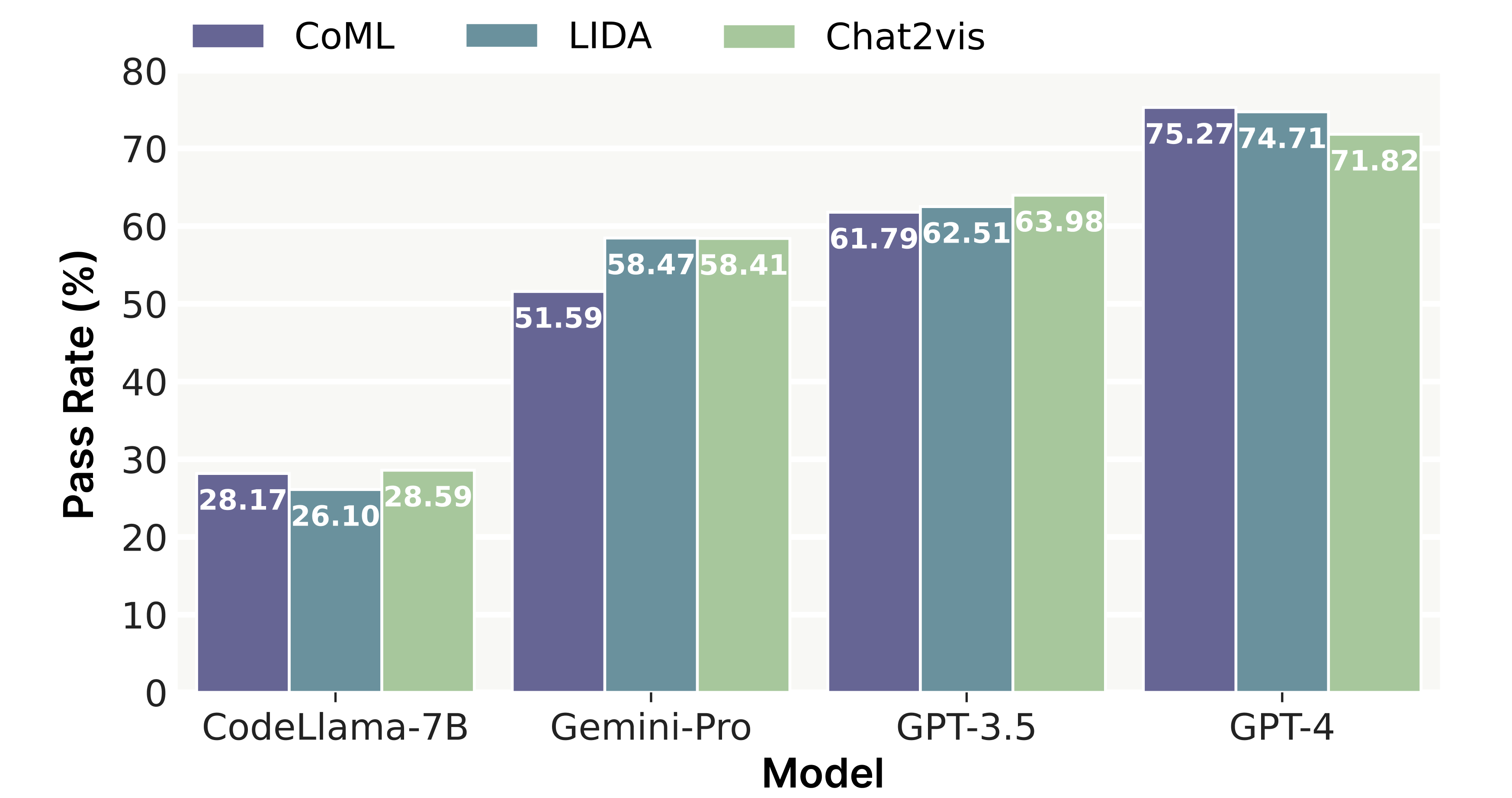}
    \vspace{-6mm}
    \caption{Comparison of pass rate across different models and table format using Matplotlib library.}
    \label{fig:format}
    \vspace{-5mm}
\end{figure}

However, different data formats have different impacts on the performance of different models. As depicted in~\cref{fig:format}, the pass rates of different models vary when generating visualizations using Matplotlib with different data formats.
We observed that when generating with GPT-3.5, using the table format of Chat2vis results in the highest pass rate.  In contrast, when generating with GPT-4, using the table format of Chat2vis yields the lowest pass rate.
These observations suggest that different LLMs exhibit preferences for specific table formats, which may stem from the use of distinct training data during pretraining. This underscores the importance of carefully selecting table formats based on the chosen LLMs.

\subsection{Table Disruption}
We conducted experiments to assess the impact of including additional unused tables when generating visualization using Matplotlib.
In our dataset, each visualization corresponds to a database that contains multiple tables but only some of these tables are used for visualization.
In this experiment, we \revise{randomly included two} unused tables in the prompt. 
If the number of unused tables is less than two, then all unused tables are added to the prompt.
As shown in~\autoref{tab:compare_table_number}, the pass rate of each LLM decreases to varying degrees. 
This suggests that it may be necessary to carefully select the required tables at the beginning of the workflow for generating visualizations.

\input{tables/compare_table_number}

%% file: tables/comparision_llms.tex
\begin{table}[t]
\centering
\small
\caption{Performance of LLMs on \name benchmark. We compare invalid rate, illegal rate, pass rate, readability score, and quality score.}
\vspace{-2mm}
\resizebox{\linewidth}{!}{%
\begin{tabular}{cc|cc|c|c|c}
\toprule
Model & Library &  \begin{tabular}{@{}c@{}}Invalid \\ Rate \end{tabular}  & \begin{tabular}{@{}c@{}}Illegal \\ Rate\end{tabular} & \begin{tabular}{@{}c@{}}Pass \\ Rate \end{tabular} & \begin{tabular} {@{}c@{}}Readability \\ Score \end{tabular}   & \begin{tabular}{@{}c@{}}Quality \\ Score \end{tabular} \\
\midrule
CodeLlama-7B & \multirow{4}*{Matplotlib}  & {\cellcolor[HTML]{e4a7a7}} \color[HTML]{000000} 42.95\%  & {\cellcolor[HTML]{edc4c4}} \color[HTML]{000000} 28.88\%  & {\cellcolor[HTML]{c4e0c7}} \color[HTML]{000000} 28.17\%  & {\cellcolor[HTML]{64ae6d}} \color[HTML]{ffffff} 3.87  & {\cellcolor[HTML]{cce4cf}} \color[HTML]{000000} 1.11 \\Gemini-Pro &   & {\cellcolor[HTML]{f6e2e2}} \color[HTML]{000000} 14.35\%  & {\cellcolor[HTML]{e9b9b9}} \color[HTML]{000000} 34.06\%  & {\cellcolor[HTML]{93c799}} \color[HTML]{000000} 51.59\%  & {\cellcolor[HTML]{429d4d}} \color[HTML]{ffffff} 3.95  & {\cellcolor[HTML]{7dbb84}} \color[HTML]{000000} 2.06 \\GPT-3.5 &   & {\cellcolor[HTML]{f9eded}} \color[HTML]{000000} 8.79\%  & {\cellcolor[HTML]{ecc2c2}} \color[HTML]{000000} 29.42\%  & {\cellcolor[HTML]{7ebb85}} \color[HTML]{000000} 61.79\%  & {\cellcolor[HTML]{f7fbf7}} \color[HTML]{000000} 3.52  & {\cellcolor[HTML]{70b478}} \color[HTML]{000000} 2.21 \\GPT-4 &   & {\cellcolor[HTML]{fdf9f9}} \color[HTML]{000000} 3.29\%  & {\cellcolor[HTML]{f1d3d3}} \color[HTML]{000000} 21.44\%  & {\cellcolor[HTML]{62ad6b}} \color[HTML]{ffffff} \bfseries 75.27\%  & {\cellcolor[HTML]{82be89}} \color[HTML]{000000} 3.80  & {\cellcolor[HTML]{379742}} \color[HTML]{ffffff} \bfseries 2.89 \\
\hline
CodeLlama-7B & \multirow{4}*{Seaborn}  & {\cellcolor[HTML]{d98585}} \color[HTML]{ffffff} 59.26\%  & {\cellcolor[HTML]{efcdcd}} \color[HTML]{000000} 24.25\%  & {\cellcolor[HTML]{ddeddf}} \color[HTML]{000000} 16.49\%  & {\cellcolor[HTML]{c5e1c8}} \color[HTML]{000000} 3.64  & {\cellcolor[HTML]{f6faf7}} \color[HTML]{000000} 0.61 \\Gemini-Pro &   & {\cellcolor[HTML]{f2d4d4}} \color[HTML]{000000} 21.09\%  & {\cellcolor[HTML]{eec8c8}} \color[HTML]{000000} 26.82\%  & {\cellcolor[HTML]{92c698}} \color[HTML]{000000} 52.09\%  & {\cellcolor[HTML]{60ac69}} \color[HTML]{ffffff} 3.88  & {\cellcolor[HTML]{7dbb84}} \color[HTML]{000000} 2.06 \\GPT-3.5 &   & {\cellcolor[HTML]{f9ecec}} \color[HTML]{000000} 9.21\%  & {\cellcolor[HTML]{ebbfbf}} \color[HTML]{000000} 31.00\%  & {\cellcolor[HTML]{82be89}} \color[HTML]{000000} \bfseries 59.79\%  & {\cellcolor[HTML]{d5e9d8}} \color[HTML]{000000} 3.60  & {\cellcolor[HTML]{70b579}} \color[HTML]{000000} 2.20 \\GPT-4 &   & {\cellcolor[HTML]{efcaca}} \color[HTML]{000000} 25.41\%  & {\cellcolor[HTML]{f5dfdf}} \color[HTML]{000000} 15.89\%  & {\cellcolor[HTML]{84bf8b}} \color[HTML]{000000} 58.70\%  & {\cellcolor[HTML]{64ae6d}} \color[HTML]{ffffff} 3.87  & {\cellcolor[HTML]{67b070}} \color[HTML]{000000} \bfseries 2.31 \\
\bottomrule
\end{tabular}%
}

\label{tab:compare_models}
\end{table}

%% file: tables/comparision_approach.tex
\begin{table}[t]
\centering
\small
\caption{A comparison of pass rate and readability scores across different approaches for queries involving a single table. The evaluation results are obtained using the GPT-3.5 model.}
\vspace{-2mm}
\resizebox{\linewidth}{!}{%
\begin{tabular}{cc|cc|c|c|c}
\toprule
Model & Library &  \begin{tabular}{@{}c@{}}Invalid \\ Rate \end{tabular}  & \begin{tabular}{@{}c@{}}Illegal \\ Rate \end{tabular} & \begin{tabular}{@{}c@{}}Pass \\ Rate \end{tabular} & \begin{tabular} {@{}c@{}}Readability \\ Score \end{tabular}   & \begin{tabular}{@{}c@{}}Quality \\ Score \end{tabular} \\
\midrule
CoML4VIS & \multirow{3}*{Matplotlib} & {\cellcolor[HTML]{f9ebeb}} \color[HTML]{000000} 4.95\%  & {\cellcolor[HTML]{dc8e8e}} \color[HTML]{000000} 27.50\%  & {\cellcolor[HTML]{91c697}} \color[HTML]{000000} 67.55\%  & {\cellcolor[HTML]{7dbb84}} \color[HTML]{000000} \textbf{3.54}  & {\cellcolor[HTML]{8bc291}} \color[HTML]{000000} \bfseries2.43 \\LIDA &   & {\cellcolor[HTML]{f7e4e4}} \color[HTML]{000000} 6.59\%  & {\cellcolor[HTML]{de9595}} \color[HTML]{000000} 25.64\%  & {\cellcolor[HTML]{90c596}} \color[HTML]{000000} 67.77\%  & {\cellcolor[HTML]{dbecdd}} \color[HTML]{000000} 2.79  & {\cellcolor[HTML]{cde5d0}} \color[HTML]{000000} 1.90 \\Chat2vis &   & {\cellcolor[HTML]{f9ebeb}} \color[HTML]{000000} 4.92\%  & {\cellcolor[HTML]{df9898}} \color[HTML]{000000} 25.15\%  & {\cellcolor[HTML]{82be89}} \color[HTML]{000000} \bfseries69.93\%  & {\cellcolor[HTML]{b4d8b8}} \color[HTML]{000000} 3.10  & {\cellcolor[HTML]{abd3b0}} \color[HTML]{000000} 2.17 \\
\hline
CoML4VIS & \multirow{3}*{Seaborn} & {\cellcolor[HTML]{f6e3e3}} \color[HTML]{000000} 6.95\%  & {\cellcolor[HTML]{d67b7b}} \color[HTML]{ffffff} 32.02\%  & {\cellcolor[HTML]{badbbe}} \color[HTML]{000000} \bfseries61.03\%  & {\cellcolor[HTML]{6db375}} \color[HTML]{000000} \textbf{3.66}  & {\cellcolor[HTML]{9ecda4}} \color[HTML]{000000} \bfseries2.27 \\
LIDA &   & {\cellcolor[HTML]{efcbcb}} \color[HTML]{000000} 12.69\%  & {\cellcolor[HTML]{d26d6d}} \color[HTML]{ffffff} 35.54\%  & {\cellcolor[HTML]{f4f9f5}} \color[HTML]{000000} 51.77\%  & {\cellcolor[HTML]{90c596}} \color[HTML]{000000} 3.39  & {\cellcolor[HTML]{ddeddf}} \color[HTML]{000000} 1.77 \\Chat2vis &   & {\cellcolor[HTML]{faf0f0}} \color[HTML]{000000} 3.65\%  & {\cellcolor[HTML]{d77f7f}} \color[HTML]{ffffff} 31.05\%  & {\cellcolor[HTML]{9fcda5}} \color[HTML]{000000} 65.30\%  & {\cellcolor[HTML]{bcdcc0}} \color[HTML]{000000} 3.04  & {\cellcolor[HTML]{beddc2}} \color[HTML]{000000} 2.02 \\

\bottomrule
\end{tabular}%
}

\label{tab:compare_approach}
\end{table}

%% file: tables/compare_table_number.tex


\begin{table}[h]
\centering
\small
\caption{Evaluating the impact of table disruption on pass rate (\%).}
\vspace{-2mm}
\begin{tabular}{c|cccc}
\toprule
Choice & CodeLlama-7B & Gemini-Pro & GPT-3.5 & GPT-4  \\
\midrule
w/o disruption & 28.17 & 51.59 & 61.69 & 75.27 \\
disruption & 17.44 \red{-10.73} & 31.80 \red{-19.79} & 54.68 \red{-7.01} & 65.86 \red{-9.41}\\

\bottomrule
\end{tabular}%
\vspace{-4mm}
\label{tab:compare_table_number}
\end{table}

%% file: sections/06-discussion.tex
\section{Discussion}

\subsection{Potential Development in NL2VIS}
The evaluation outcomes in Section~\ref{sec:evaluation} show that current methods in NL2VIS still have room for improvement. 
These highlight the importance of enhancing the performance of LLMs by exploring advanced techniques and knowledge in natural language processing and data visualization.
We discuss the potential development as follows:

\begin{itemize}[noitemsep,topsep=0pt,leftmargin=10pt]
    \item Incorporating supplementary methods, such as linting methods like pylint~\cite{pylint}, can help address issues such as omissions in package imports. Linting methods analyze code for errors and style inconsistencies, providing proactive guidance and enhancing code quality.
    \item Observing frequent misuses of APIs by LLMs, it is essential to develop strategies that guide LLMs using library API documentation to improve the accuracy of API usage. Such strategies may involve techniques such as retrieval augmented generation (RAG)~\cite{lewis2020retrieval} or model fine-tuning.
    \item Decomposing NL2VIS tasks into subtasks is another effective approach, addressing errors across multiple steps from data transformation to visualization transformation.
    Therefore, simplifying complex problems into manageable steps like data understanding, column selection, visual mapping, and sorting can lead to more accurate and efficient results.
    \item Iterative generation guided by feedback to refine and improve the quality of generated visualizations.
    While it is challenging to directly detect issues through code alone, integrating visual-based methods can provide valuable feedback. For instance, incorporating our readability evaluator into the generation process helps identify readability issues in the generated results, guiding subsequent modifications. 
\end{itemize}

\subsection{Limitations and Future work}
We summarize several limitations and propose future work direction.

\revise{
\underline{\textit{Support for integrating additional grammar or methods.}}
In this work, we evaluate visualizations generated using Python libraries. However, our framework's modular design makes it easy to extend and evaluate other visualization generation tools, such as JavaScript-based toolkits.
To evaluate Vega-Lite-based methods, for instance, we need to configure the code execution and construction module. Specifically, we simulate a browser environment to convert Vega-Lite code into SVG format and adapt our deconstruction rules to accurately extract data from the rendered charts.
As a result, our framework is capable of adaptively and robustly evaluating a broad spectrum of automatic visualization tools. This level of flexibility ensures that our framework remains useful and adaptable to emerging visualization technologies.
}

\revise{
\underline{\textit{Expand the scope of benchmark.}}
Currently, our dataset focuses on common chart types. While it provides a solid foundation for benchmarking, it does not encompass the full range of natural language queries and visualizations. In the future, we aim to create a more comprehensive and challenging benchmark that can drive further advancements in the field of NL2VIS. We plan to collaborate with BI-tools teams and the broader community to expand our benchmark, including real-world queries and more complex visualizations. By leveraging our proposed construction process, which integrates the capabilities of state-of-the-art LLMs with insights from human experts, we anticipate enhancing the efficiency and quality of future dataset expansions.}

\underline{\textit{Extend the coverage of metrics.}}
The evaluation dimensions covered in \name primarily focus on fundamental errors that hinder comprehension. 
At this point, we have not included metrics related to aesthetics or expressiveness. 
This is partly because these aspects represent higher-level requirements that are not the primary challenges currently faced. 
Additionally, they involve more subjective considerations that are influenced by the visualization's intended use and audience, necessitating more complex evaluations. 
\revise{In the future, we plan to expand the assessments to include aesthetic, expressiveness, and stylistic aspects by leveraging more advanced models, enhancing the performance and capability of \name.}



%% file: sections/07-conclusion.tex
\section{Conclusion}

We present \name, a novel NL2VIS benchmark aimed at comprehensive and reliable evaluation of generated visualizations.
Our work includes the construction of a large-scale and high-quality dataset, the development of an automated evaluation framework covering dimensions of validity, legality, and readability, and the evaluation of state-of-the-art LLMs.
Our evaluations reveal common challenges of LLMs, offering valuable insights for future advancements.
Overall, our framework represents a significant step forward in improving the quality of NL2VIS systems in the era of LLMs.

%% file: sections/appendix.tex
\setcounter{figure}{0} 
\renewcommand\thefigure{\Alph{figure}} 

\setcounter{table}{0}
\renewcommand\thetable{\Alph{table}}

\begin{appendices}
\section{Dataset Construction Details}
\label{sec:dataset_construction_details}
\subsection{\revise{Issues in nvBench}}
\revise{The queries from nvBench exhibit several issues, including  \textbf{irrationality}, \textbf{ambiguity}, \textbf{duplication}, and \textbf{erroneous labels}.
Table~\ref{tab:benchmark-shortcomings} presents examples of these issues along with explanations.}
\input{tables/nvbench_issues}

\subsection{\revise{Existing Datasets}}
\revise{As shown in Table~\ref{tab:compare_existing_dataset}, previous visualization datasets either focus solely on narrow domains or lack accurate pairs of natural language and visualizations.}
\input{tables/compare_dataset}

\subsection{High-quality Queries Selection}
\revise{High-quality query selection, considering both reliability and cost, involves three steps: rule-based, LLMs-based, and human-based selection. Rule-based methods are the most economical and convenient, but they are also limited to filtering out low-quality queries with easily describable characteristics. Human-based selection is the most reliable but also the most labor-intensive. LLMs-based methods strike a balance, offering greater flexibility than rule-based methods while being more convenient than human-based approaches. Therefore, we first use rule-based methods to eliminate queries with clear, identifiable shortcomings. Next, LLMs-based methods provide further filtering. Finally, humans review the remaining smaller dataset to ensure high reliability in a cost-effective manner.}

\subsubsection{Rule-based Queries Selections}
We filtered 5,343 visualizations (\textit{VIS}) and 19,898 natural language and visualization (\textit{NL}, \textit{VIS}) pairs through rule-based selection. 
\revise{These rules can be efficiently implemented using simple expressions, such as regular expressions, to identify or correct clear, identifiable issues. 
While we acknowledge that these rules may not completely eliminate all low-quality queries, they serve as the first step to reduce the workload for subsequent LLM and human evaluations.
The selection rules are listed below, along with the number of \textit{VIS} and (\textit{NL}, \textit{VIS}) pairs filtered or corrected.}


\revise{To ensure that the generated visualizations are based on \textbf{valid data}, we apply rule R1 to filter out cases where the original data is empty. }

\noindent\textbf{R1}: Remove (\textit{NL}, \textit{VIS}) pairs that require a table, but the table is empty. (\minus{\textit{-15 VIS, -15 (NL, VIS)}})

Examples:
\begin{itemize}[noitemsep,topsep=0pt]
    \item {empty table ``staff'' (\textit{2687})}     
    \item {empty table ``players'' (\textit{3286})} 
\end{itemize}

\revise{To guarantee the \textbf{rationality}, we introduce rules R2 and R3, which remove visualizations where identical data is mistakenly used as numerical data and where irrational ordering of temporal or nominal data is present. }

\noindent\textbf{R2}: Remove (\textit{NL}, \textit{VIS}) pairs where identified data such as ID or code are treated as numerical. (\minus{\textit{-1394 VIS, -6442 (NL, VIS)}})

Examples:
\begin{itemize}[noitemsep,topsep=0pt]
    \item {``Show the \error{faculty id} of each faculty member, along with the number of students he or she advises in a scatter chart'' (\textit{3})}     
    \item {``Show me a scatter plot of \error{code} and minimal price for.'' (\textit{2184})} 
\end{itemize}

\noindent\textbf{R3}: Remove (\textit{NL}, \textit{VIS}) pairs that have inappropriate sorting requirements ``from high to low'' or ``from low to high'' for temporal or nominal data. (\minus{\textit{-685 VIS, -2159 (NL, VIS)}})

Examples:
\begin{itemize}[noitemsep,topsep=0pt]
    \item {``Show all dates of transactions whose type code is "SALE", and count them by a line chart, and list x axis \error{from high to low order}.'' (\textit{2990@x\_name@DESC})}     
    \item {``Bar chart x axis location y axis sum total passengers, show from \error{low to high by the x axis}." (\textit{3053@x\_name@ASC})} 
\end{itemize}

\revise{To eliminate \textbf{ambiguity} in the queries, we use rules R4 and R5 to filter out vague sorting or to correct ambiguous binning. }

\noindent\textbf{R4}: Remove (\textit{NL}, \textit{VIS}) pairs that have ambiguous sorting requirements, such as ``sort bars in desc order'', ``order by the bar in descending'', ``rank by the bars in asc''. (\minus{\textit{-107 VIS, -583 (NL, VIS)}})

Examples:
\begin{itemize}[noitemsep,topsep=0pt]
    \item {``A bar chart listing the number of faults for different description of skills required to fix them, \error{show by the bar in asc}.'' (\textit{149@x\_name@AS})}     
    \item {``Group and count brand for each camera lens using a bar chart, \error{sort by the bar in ascending}.'' (\textit{2349@x\_name@ASC})} 
\end{itemize}

\noindent\textbf{\revise{R5}}: Rewrite \textit{NL} that have ambiguous binning requirements such as ``by time''. The ground truth explicitly includes binning by year, month, or weekday. Replace the ambiguous term ``by time'' with a clearer bin granularity.(\minus{\textit{63 VIS, 313 (NL, VIS)}})

Examples:
\begin{itemize}[noitemsep,topsep=0pt]
    \item ``For those employees who did not have any job in the past, show me about the distribution of  hire\_date and the sum of salary bin hire\_date \error{by time} in a bar chart.'' (\textit{1716})  
    \item {``Visualize a line chart about the change of the amount of Start\_from over Start\_from bin start\_from \error{by time}, order X-axis in descending order.'' (\textit{1356@x\_name@DESC })} 
\end{itemize}

\revise{For \textbf{erroneous labels}, rule R6 helps in removing cases where the ground truth contains incorrect binning data.} 

\noindent\textbf{\revise{R6}}: Remove (\textit{NL}, \textit{VIS}) pairs with incorrect binning data, specifically, cases where the query specifies binning by year or does not specify binning, but the ground truth combines multiple consecutive years into one bin. (\minus{\textit{-123 VIS, -871 (NL, VIS)}})

Examples:
\begin{itemize}[noitemsep,topsep=0pt]
    \item {``\error{Bin the transcript date into year interval} and count them for a line chart, list from low to high by the X.'' (\textit{2891@x\_name@ASC})}     
    \item {``What are the number of dates of birth of all the guests whose gender is "Male"?'' (\textit{80})} 
\end{itemize}

\revise{For \textbf{duplication} issues, we apply rule R7 to randomly remove similar queries. }

\noindent\textbf{\revise{R7}}: Randomly remove similar \textit{VIS} that have almost identical descriptions, except for different sorting requirements. These \textit{VIS} produce different results by sorting the x-axis and y-axis differently. However, due to the high similarity in NL queries, repeating these executions adds little value but incurs significant resource overhead. Therefore, we randomly select one \textit{VIS} from the same prefix IDs to reduce redundancy. (\minus{\textit{-3535 VIS, -10354 (NL, VIS)}})

Examples:
\begin{itemize}[noitemsep,topsep=0pt]
    \item {``\error{Draw a bar chart of operating system versus the total number}''(\textit{372}), ``\error{Draw a bar chart of operating system versus the total number}, list from low to high by the X-axis.''(\textit{372@x\_name@ASC}), ``\error{Draw a bar chart of operating system versus the total number}, display from high to low by the names please.''(\textit{372@x\_name@DESC}), `\error{Draw a bar chart of operating system versus the total number}, order in asc by the Y.''(\textit{372@y\_name@ASC}), ``\error{Draw a bar chart of operating system versus the total number}, and list by the Y in desc please.''(\textit{372@y\_name@DESC}).}     
    \item {``\error{Find the number of web accelerators used for each Operating system}.''(\textit{372}), ``\error{Find the number of web accelerators used for each Operating system}, I want to display in ascending by the X.''(\textit{372@x\_name@ASC}), ``\error{Find the number of web accelerators used for each Operating system}, and I want to rank x-axis from high to low order.''(\textit{372@x\_name@DESC}), `\error{Find the number of web accelerators used for each Operating system}, I want to display by the y-axis in ascending.''(\textit{372@y\_name@ASC}), ``\error{Find the number of web accelerators used for each Operating system}, show from high to low by the total number.''(\textit{372@y\_name@DESC}).}  
\end{itemize}

\revise{Furthermore, for \textbf{out-of-scope} histograms, we use rule R8 to remove them.} 

\noindent\textbf{\revise{R8}}: Remove (\textit{NL}, \textit{VIS}) pairs that involve drawing histograms, as our evaluation framework plans to support this chart type in the future. (\minus{\textit{-3 VIS, -6 (NL, VIS)}})

Examples:
\begin{itemize}[noitemsep,topsep=0pt]
    \item {``Show me how many long by long in a \error{histogram}, could you rank from low to high by the y axis?'' (\textit{327@y\_name@ASC})} 
    \item {``For each station, \error{bin} its longitude \error{divided by zero as buckets}, and count the frequency in each bucket.'' (\textit{327})}     
    
\end{itemize}

\subsubsection{LLMs-based Queries Selections}
\begin{figure}[H]
    \centering    \includegraphics[width=\linewidth]{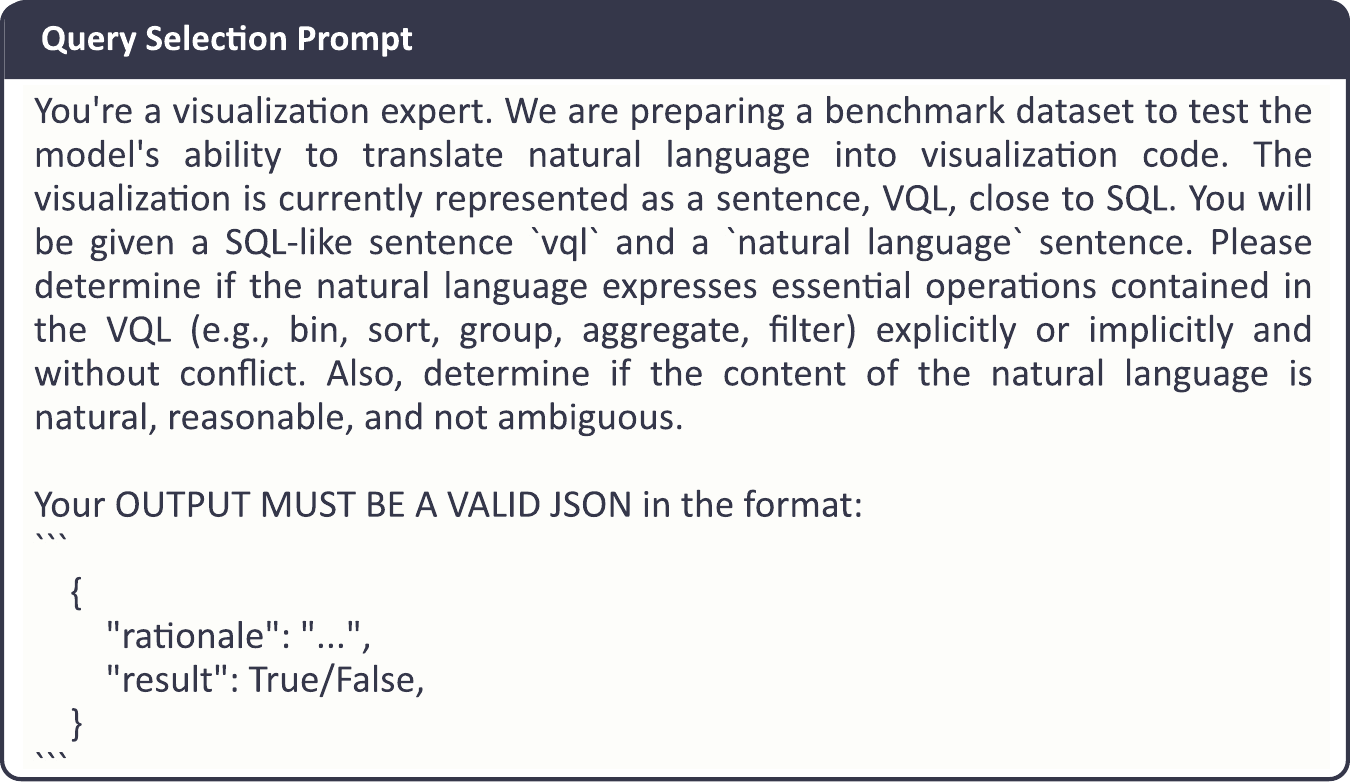}
    \vspace{-3mm}
    \caption{The prompt of selecting appropriate queries. }
    \label{fig:voting_prompt}
    \vspace{1mm}
\end{figure}

We employed three LLMs (i.e., GPT-4, GPT-3.5, Gemini-Pro) to verify whether the natural language specifies all essential operations contained in ``VQL'' and is reasonable and unambiguous. ``VQL'' is a SQL-like sentence describing the visualization type and details of data transformation. The prompt is shown in \cref{fig:voting_prompt}. 
Additionally, We adopted a few-shot prompt strategy, selecting typical examples to facilitate the LLMs' understanding of the task.
In total, we filtered 255 \textit{VIS} and 1,432 (\textit{NL}, \textit{VIS}) pairs through LLMs-based selection.

\subsubsection{Human Selections}
Human experts are involved in ensuring the quality of queries by reviewing and handling them according to the following principle: if there are multiple other queries in the dataset that use the same database and represent the same chart type as the query under review, it is deleted directly; otherwise, it is manually modified or rewritten to create a new query. 
Additionally, queries corresponding to chart types with a low number of instances are supplemented as needed. 
303 \textit{VIS} and 1,369 (\textit{NL}, \textit{VIS}) pairs are filtered by human. 
We list some filtered examples as follows.

\textbf{Ambiguous Examples:} 
\begin{itemize}[noitemsep,topsep=0pt]
    \item {\textit{Query}: ``Plot how many hire date by grouped by hire date as a bar graph''  (\textit{1839})}
    
    \textit{VQL}: ``Visualize BAR SELECT HIRE\_DATE , COUNT(HIRE\_DATE) FROM employees \error{BIN HIRE\_DATE BY WEEKDAY}'' 

    \textit{Reason}: The query does not specify bin granularity.
    
    \item {\textit{Query}:  ``What are the names of catalog entries with level number 8, and count them by a pie chart'' (\textit{2608})}
    
    \textit{VQL}: ``Visualize PIE SELECT catalog\_entry\_name , COUNT(catalog\_entry\_name) FROM \error{Catalog\_Contents AS t1 JOIN Catalog\_Contents\_Additional\_Attributes AS t2 ON t1.catalog\_entry\_id = t2.catalog\_entry\_id} WHERE t2.catalog\_level\_number = "8" GROUP BY catalog\_entry\_name'' 

    \textit{Reason}: The query does not provide an explanation for why joining Table Catalog\_Contents\_Additional\_Attributes is necessary. 
\end{itemize}

\noindent\textbf{Irrational Examples:} 
\begin{itemize}[noitemsep,topsep=0pt]
    \item {\textit{Query}: ``Find the last name and \error{age} of the student who has allergy to both milk and cat. Show a \error{pie} chart.'' (\textit{52})} 
    
    \textit{Reason}: Age is not suitable for calculating percentages and displaying in a pie chart.
    
    \item {\textit{Query}: ``Show the number of customers for each \error{gender}. Show \error{scatter chart}.'' (\textit{925})} 
    
    \textit{Reason}: Although gender is represented as 0 and 1, it is a categorical variable and not suitable for creating a scatter plot.
\end{itemize}

\noindent\textbf{Redundant Examples:} 
\begin{itemize}[noitemsep,topsep=0pt]
    \item {\textit{Filter condition}: ``WHERE salary BETWEEN 8000 AND 12000 AND commission\_pct != "null" OR department\_id != 40'' (\textit{e.g., 1581, 1584, 1609})} 
    
    \textit{Reason}: In nvBench, \error{33 VIS} were generated based on the data filtered using the aforementioned condition. We randomly removed some of them, aiming to mitigate potential bias introduced by this type of data processing.
    
    \item {\textit{Filter condition}: ``WHERE employee\_id NOT IN (SELECT employee\_id FROM job\_history)'' (\textit{e.g., 1710@y\_name@DESC, 1714@y\_name@ASC, 1722@y\_name@DESC})} 
    
    \textit{Reason}: In nvBench, \error{48 VIS} were generated based on the data filtered using the aforementioned condition. We randomly removed some of them, aiming to mitigate potential bias introduced by this type of data processing.
\end{itemize}

\subsection{Accurate Ground Truth Labeling}
\subsubsection{Meta-information}
We added three aspects of meta-information for each query to facilitate reliable comparisons.

\textbf{Specified channel} is an array that stores explicitly or implicitly specified channels.
For example, in the query ``Show the number of customer address history in each day and group by date to with a line chart.'', the channels ``x'' and ``y'' are implicitly specified. In the query ``Scatter plot to show consider rate on x axis and oppose\_rate on y axis'', the channels ``x'' and ``y'' are explicitly specified.

\textbf{Sort requirement} is a dictionary that has three keys: ``channel'', ``order'', and ``sort\_by''. The ``channel'' key specifies the visualization axis (e.g., x-axis, y-axis) to which the sorting should be applied. The ``order'' key indicates the desired order of sorting (ascending or descending), while the ``sort\_by'' key represents the criterion for sorting, i.e., the field or axis based on which the sorting should be performed.
For example, the query ``Compare the average age of drivers from the same home city with a bar chart, could you rank in descending by the Y-axis please?'' has the following sort requirement: \texttt{\{"channel":"y","order":"descending","sort\_by":"axis"\}}. Similarly, the query ``Plot the average of age by grouped by home city as a bar graph, show from high to low by the total number.'' has the following sort requirement: \texttt{\{"channel":"y","order":"descending","sort\_by":"field"\}}.

\textbf{Stacked bar} is a boolean value indicating whether the query explicitly or implicitly specifies stacked bar.
Although we filtered out queries that did not specify the chart type, there are still cases where the query mentions ``bar'', indicating that both grouping bar and stacked bar may be suitable. For example, query ``Compute the total number in each competition type, and then split by country. Plot a bar chart and show in desc by the x-axis.'' does not specify stacked bar (False). On the contrary, the query ``How many courses each teacher taught? Show me a stacked bar chart. The x-axis is the teacher's first name and group by course description.'' specifies a stacked bar (True).

\subsection{Dataset Rebalancing}
197 simple \textit{VIS} and 539 (\textit{NL}, \textit{VIS}) pairs are excluded in this step to balance the dataset and achieve a moderate level of complexity across various chart types.

\subsection{\revise{VisEval: Statistics}}
\begin{figure}[bth]
    \centering    \includegraphics[width=\linewidth]{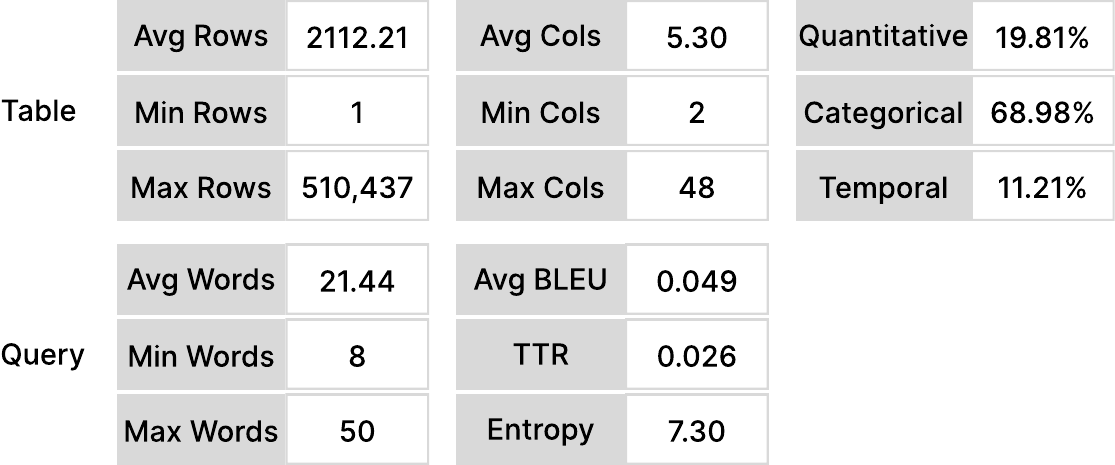}
    \vspace{-5mm}
    \caption{\revise{Statistics of table and query.}}
    \label{fig:dataset_detail}
    \vspace{-3mm}
\end{figure}

\revise{VisEval encompasses 146 databases with a total of 748 tables. As shown in~\cref{fig:dataset_detail}, the tables vary significantly in size, with the number of rows ranging from 1 to 510,437, and an average of 2,112.21 rows per table. Among the columns, 19.51\% are quantitative, 69.75\% are categorical, and 10.74\% are temporal.
}

\revise{In summary, VisEval includes 2,524 representative queries covering seven common chart types. On average, each query in VisEval contains 21.44 words, with the longest query consisting of 50 words.
To evaluate the quality and diversity of these queries, we utilized several metrics: Bilingual Evaluation Understudy (BLEU), Type-Token Ratio (TTR), and word entropy.
BLEU measures the similarity of text to a reference text, with lower scores indicating a greater difference.
TTR assesses the lexical diversity of the dataset, with higher values indicating greater diversity. 
Word entropy measures the unpredictability or variability of word usage, with higher values indicating a richer variety of words.
The average BLEU score of queries in VisEval is 0.049, the TTR is 0.026, and the word entropy is 7.30. For comparison, the average BLEU score of queries in the original nvBench dataset is 0.052, the TTR is 0.003, and the word entropy is 7.21. These metrics suggest that our dataset offers a more diverse and lexically rich set of queries compared to the original nvBench dataset.}

\revise{Furthermore, based on a low-level visual analytical task taxonomy~\cite{amar2005low}, we conducted an analysis of query types. Given the low-level nature of these visualization tasks, complex queries may encompass multiple tasks simultaneously. Table~\ref{tab:task_types} presents the distribution and proportions of different tasks within the dataset. The most prevalent task is ``Compute Derived Value'', which is fundamental and often accompanies other tasks, indicating that a majority of queries involve computing aggregations (e.g., average, median, count). Compared to straightforward ``Retrieve Value'' tasks, ``Compute Derived Value'' tasks tend to be more challenging.
Additionally, ``Sort'' tasks are also prominent, typically occurring alongside other tasks. Sorting is essential for many analytical workflows, enhancing data interpretation. The dataset also exhibits notable occurrences of tasks such as ``Filter'', ``Characterize Distribution'', ``Correlate'', and ``Find Extreme''. This diversity suggests that our dataset encompasses a broad range of visual analysis tasks, highlighting its generality and applicability.}

\input{tables/task}

\section{\revise{Evaluation framework Details}}
\subsection{\revise{Legality Checker}}
\label{sec:legality_checker}

\begin{figure*}[bth]
    \centering    \includegraphics[width=0.9\linewidth]{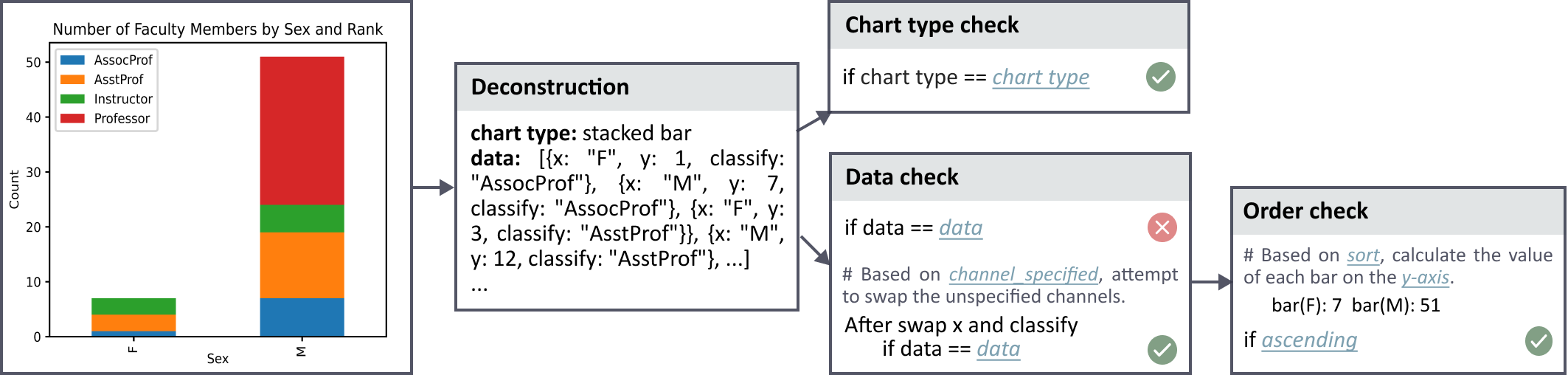}
    \vspace{0mm}
    \caption{\revise{Example of the legality check. In this instance, the generated visualization has passed the chart type check, data check, and order check, and is therefore considered legal. The queries and ground truth for this example are presented in~\cref{fig:dataset_example}. In this figure, \underline{\textit{information}} from the ground truth is underlined for emphasis.}}
    \label{fig:example_legality_check}
    \vspace{0mm}
\end{figure*}

\revise{As shown in~\cref{fig:example_legality_check}, the legality checker begins by deconstructing the SVG charts to obtain the chart type, plotted data, and visual mapping. After deconstruction, chart type check and data check are executed in parallel. The chart type check compares the extracted chart type with the ground truth; if they match, the chart type is considered correct. If there is a mismatch, the meta-information's strict\_stacked\_bar attribute is used to determine whether a strict stacked bar chart is required or if a grouped bar chart is acceptable. For the data check, the first step is to compare the extracted data with the ground truth (ignoring order). If they do not match, the method attempts to swap unspecified channels based on channel\_specified. If the data matches after swapping, it is considered to have passed the data check. Only data that passes the data check proceeds to the order check. During the order check, the values are first extracted based on the sort requirements, and it is then verified whether these values are sorted in the specified order.}

\subsection{Readability Evaluator}
\label{sec:readability_evaluator}
\begin{figure}[htb]
    \centering    \includegraphics[width=\linewidth]{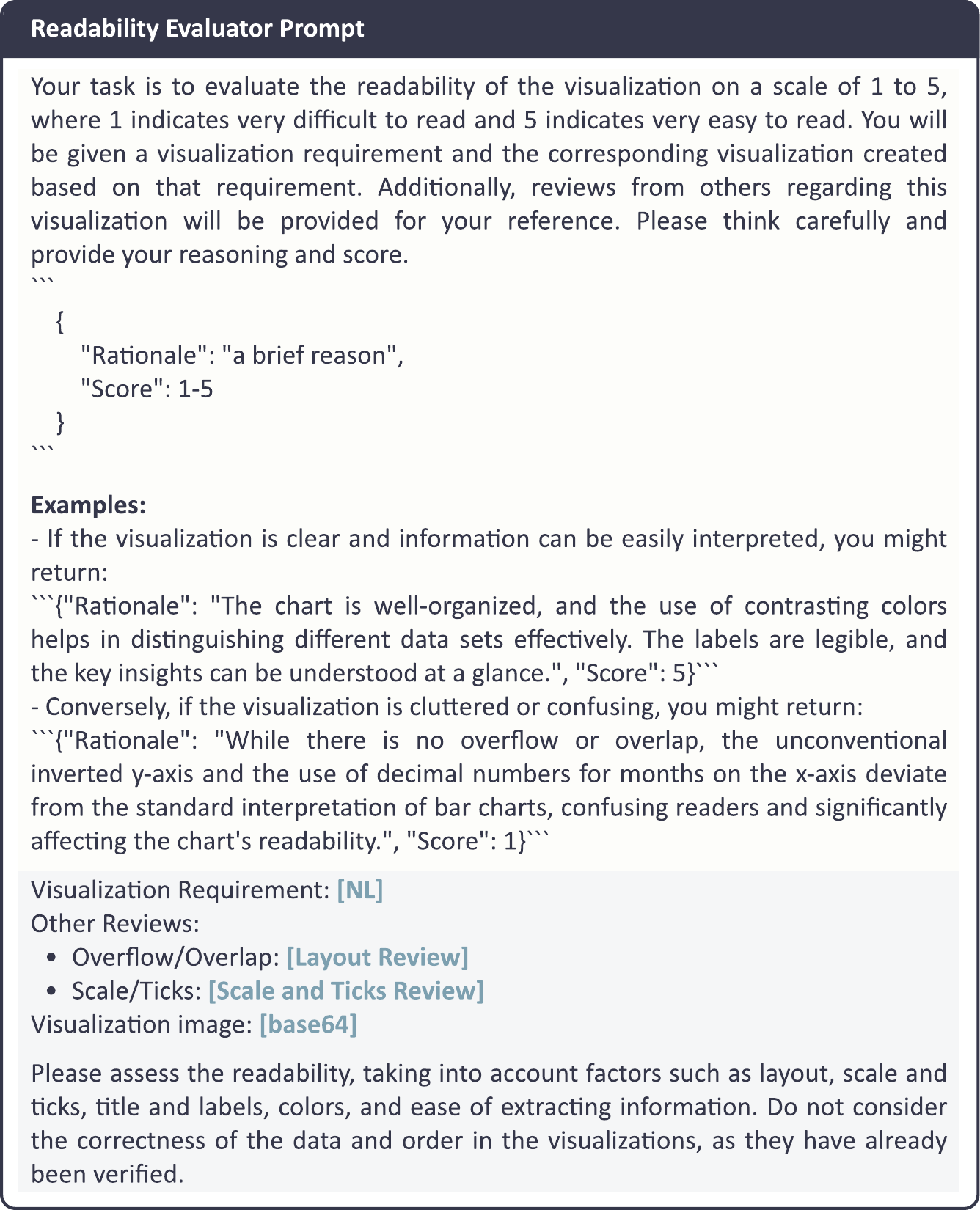}
    \vspace{-6mm}
    \caption{The prompt template for the readability evaluator. }
    \label{fig:readablity_prompt}
    \vspace{-4mm}
\end{figure}

\begin{sidewaysfigure*}[p]
    \centering
    \includegraphics[width=0.85\linewidth]{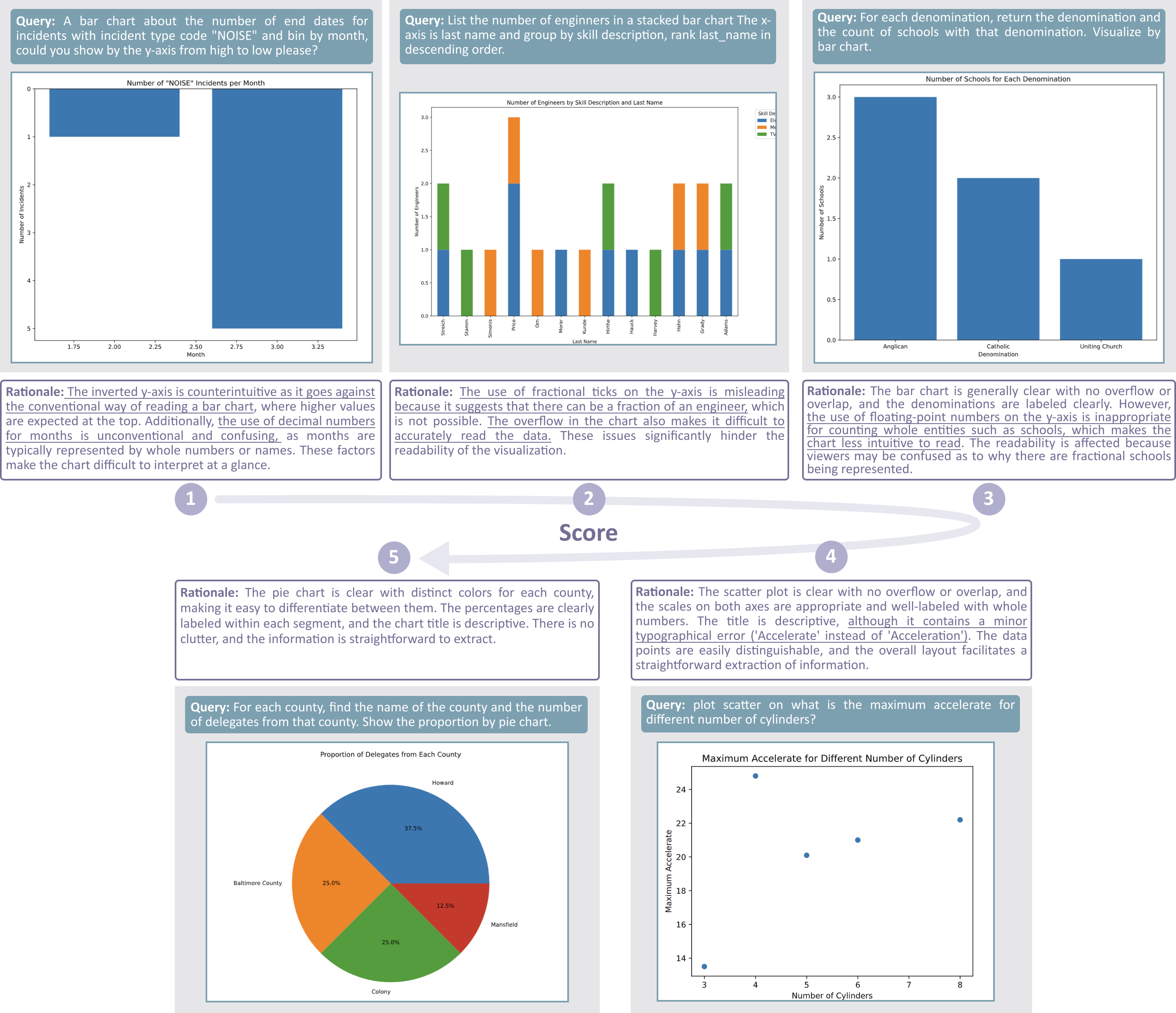}
    \caption{ \revise{Detailed readability evaluation results across scores from 1 to 5. For each example, the query, the evaluated visualization, and the rationale provided by the readability evaluator are displayed.} }
    \label{fig:score_examples}
\end{sidewaysfigure*}

The prompt template for the readability evaluator is illustrated in~\cref{fig:readablity_prompt}. 
\revise{Additionally, detailed examples across scores from 1 to 5 are presented in~\cref{fig:score_examples}.}

\subsubsection{\revise{Discussion}}
\label{sec:readability_discussion}
\revise{As shown in~\cref{fig:readability_correlation}, our method exhibits a high correlation with human ratings. However, we observed discrepancies between the ratings given by human experts and the readability evaluator. These inconsistencies are also evident among human experts themselves.
Upon further analysis, we identified two primary causes for these individual differences. Firstly, individuals have varying judgments on the impact of the same readability issues. While they may identify the same issues, their assessments of severity differ. This is due to differences in experience, perception, and personal standards. For example, one expert might consider a minor labeling issue to be significant, while another might view it as negligible.
Secondly, readability evaluation is inherently subjective and influenced by numerous factors such as fatigue, personal biases, and the limitations of the evaluation method itself. These factors can lead to the oversight or misjudgment of certain aspects. Given the subjective nature of readability assessment, such discrepancies are understandable and expected.}

\begin{figure}[htb]
    \centering    \includegraphics[width=0.75\linewidth]{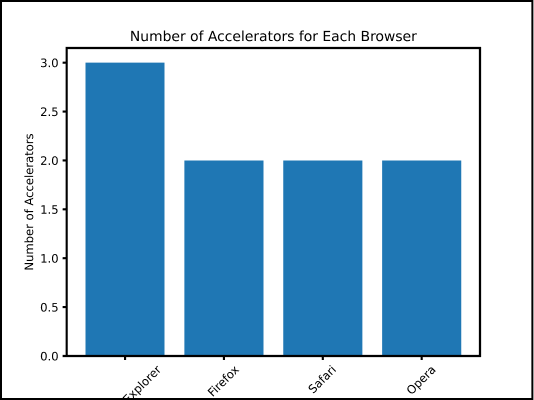}
    \vspace{-2mm}
    \caption{\revise{An example for readability evaluator.}}
    \label{fig:readablity_example}
    \vspace{-2mm}
\end{figure}

\revise{For example, in~\cref{fig:readablity_example}, the scores given by the three human experts and the readability evaluator were 2, 3, 4, and 2, respectively. 
The readability evaluator rated with a rationale ``The bar chart has an issue with the y-axis scale using floating-point numbers, which is not suitable for counting whole items like accelerators. Additionally, there is an overflow problem which could obscure important data. These issues can lead to confusion and misinterpretation of the data presented.''
Through interviews, we found that Expert 1, who rated the same score as the GPT-4V evaluator, agreed that the ticks and overflow issues significantly impacted readability. He mentioned that the floating-point numbers on the y-axis were confusing for representing whole items, and the overflow issue made it difficult to interpret the data correctly.
Expert 2 had observed both the ticks and overflow issues but rated the chart a 3. She explained that although the text overflowed the canvas, it was only slightly, and she still found the chart understandable. Her slightly higher rating reflects her tolerance for minor readability issues that do not completely obscure the information.
Expert 3, who gave the highest rating of 4, indicated that he only noticed the overflow issue during his evaluation. However, he also acknowledged that using floating-point numbers for whole item counts on the y-axis was not ideal.
These examples highlight the subjective nature of readability evaluations and the varying thresholds of different experts.
}

\revise{We acknowledge the inherent subjectivity in readability evaluations and recognize that using a consistent model for evaluating visualizations can help reduce the score inconsistencies caused by individual preferences. Additionally, we anticipate that advancements in evaluation models will further minimize errors in judgment over time, enhancing the reliability and accuracy of readability assessments.}

\section{Supplementary Evaluations}
\label{sec:evaluation_choices}
\subsection{Few-shot Prompting}
\label{sec:one-shot}
\begin{figure}[ht]
    \centering    \includegraphics[width=\linewidth]{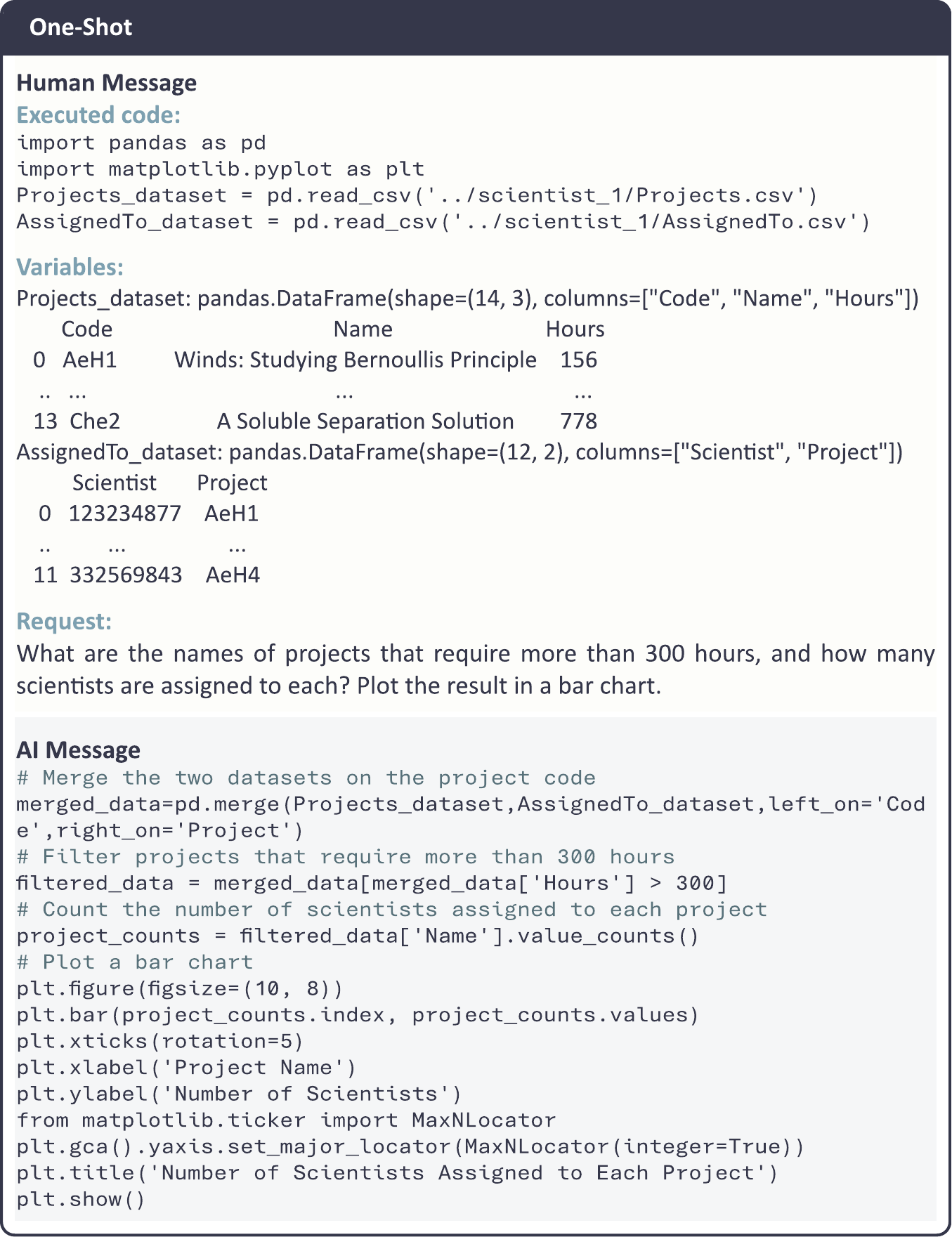}
    \vspace{-5mm}
    \caption{The one shot for CoML4VIS. }
    \label{fig:one_shot}
    \vspace{-2mm}
\end{figure}

Few-shot prompting is a technique that leverages examples provided in prompts to aid the model in understanding the context better.
By default, CoML4VIS employs a 1-shot prompt, which includes one example.
We meticulously selected a bar chart instance that combines data from two tables as an example. In the example code, we rotated and adjusted the ticks to display integers, thereby enhancing the chart's readability, as shown in~\cref{fig:one_shot}.

\begin{figure}[ht]
    \centering    \includegraphics[width=\linewidth]{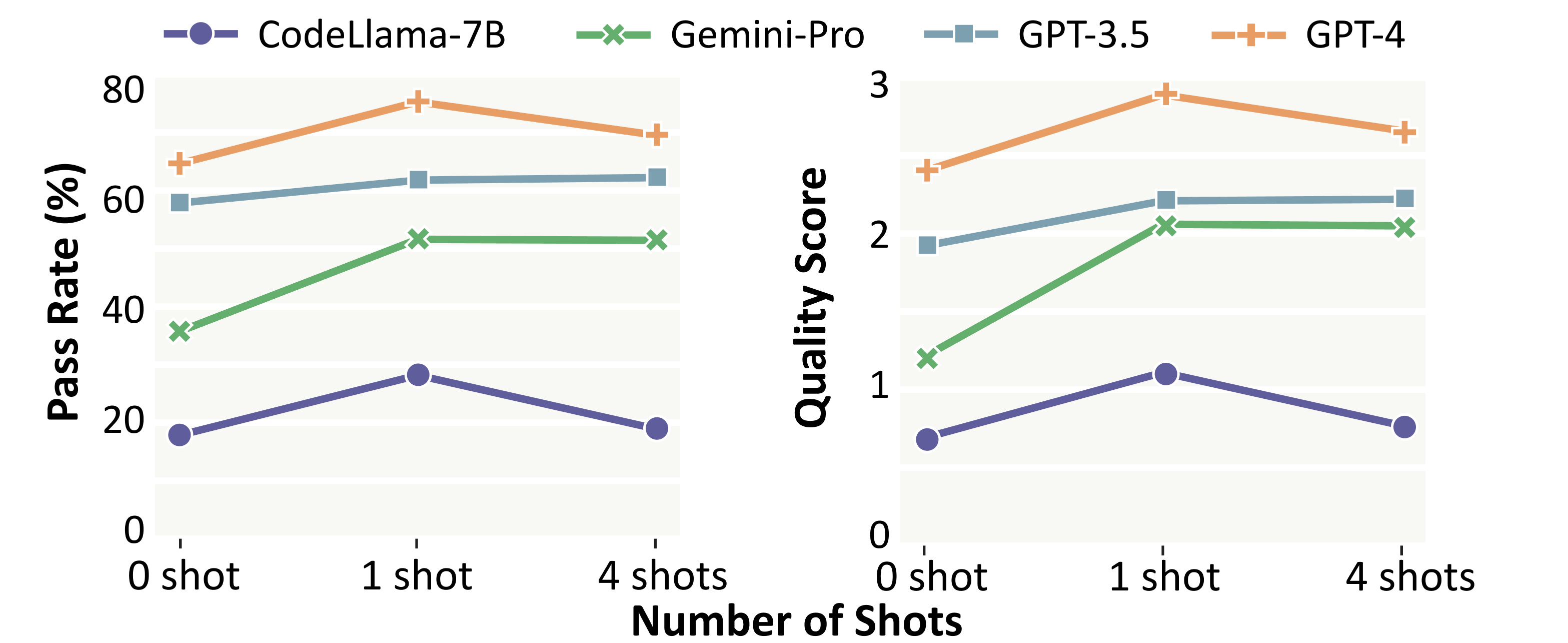}
    \vspace{-5mm}
    \caption{Impact of numbers of shots used for few-shot prompting.}
    \label{fig:fewshot}
    \vspace{-5mm}
\end{figure}

We also conducted experiments using the zero-shot setting and 4-shot setting for comparison. In the 4-shot setting, we used the example from the one-shot setting as the first one, and we also included examples of a grouping scatter chart, a grouping line chart, and a pie chart to diversify the examples.
As shown in~\cref{fig:fewshot}, there is a significant decrease in both the pass rate and quality score in a zero-shot setting compared to one-shot setting. 
However, most models do not achieve better performance in 4-shot setting with richer context. This is possibly due to the increased length of the input text causing the models to lose focus~\cite{liu2024lost}. 
Further ablation is needed to understand the underlying cause of this gap.

\subsection{\revise{Evaluating Different Number of Rows}}
\revise{By default, when sampling rows to describe the table, we limited the number to 10 in our evaluation. We further conducted tests to assess whether there is a significant difference in the pass rate with an increased number of rows when generating visualizations using GPT-3.5 on the Matplotlib library. As shown in Table~\ref{tab:compare_number_rows}, the pass rate did not show any significant differences. Therefore, we believe that limiting the number of rows to 10 during evaluation is a reasonable choice, as it effectively demonstrates the model's capabilities while consuming fewer tokens.}

\begin{table}[ht]
\centering
\caption{\revise{Impact of the number of rows used to describe the table.}}
\vspace{-2mm}
\begin{tabular}{c|ccc}
\toprule
 Number of Rows & 10 & 20 & 30  \\
\midrule
Pass Rate & 61.79\% & 61.50\% & 61.68\% \\
\bottomrule
\end{tabular}
\label{tab:compare_number_rows}
\end{table}

\subsection{Typical Errors}
\label{sec:typical_errors}

\begin{figure*}[t!]
    \centering    \includegraphics[height=0.95\textheight]{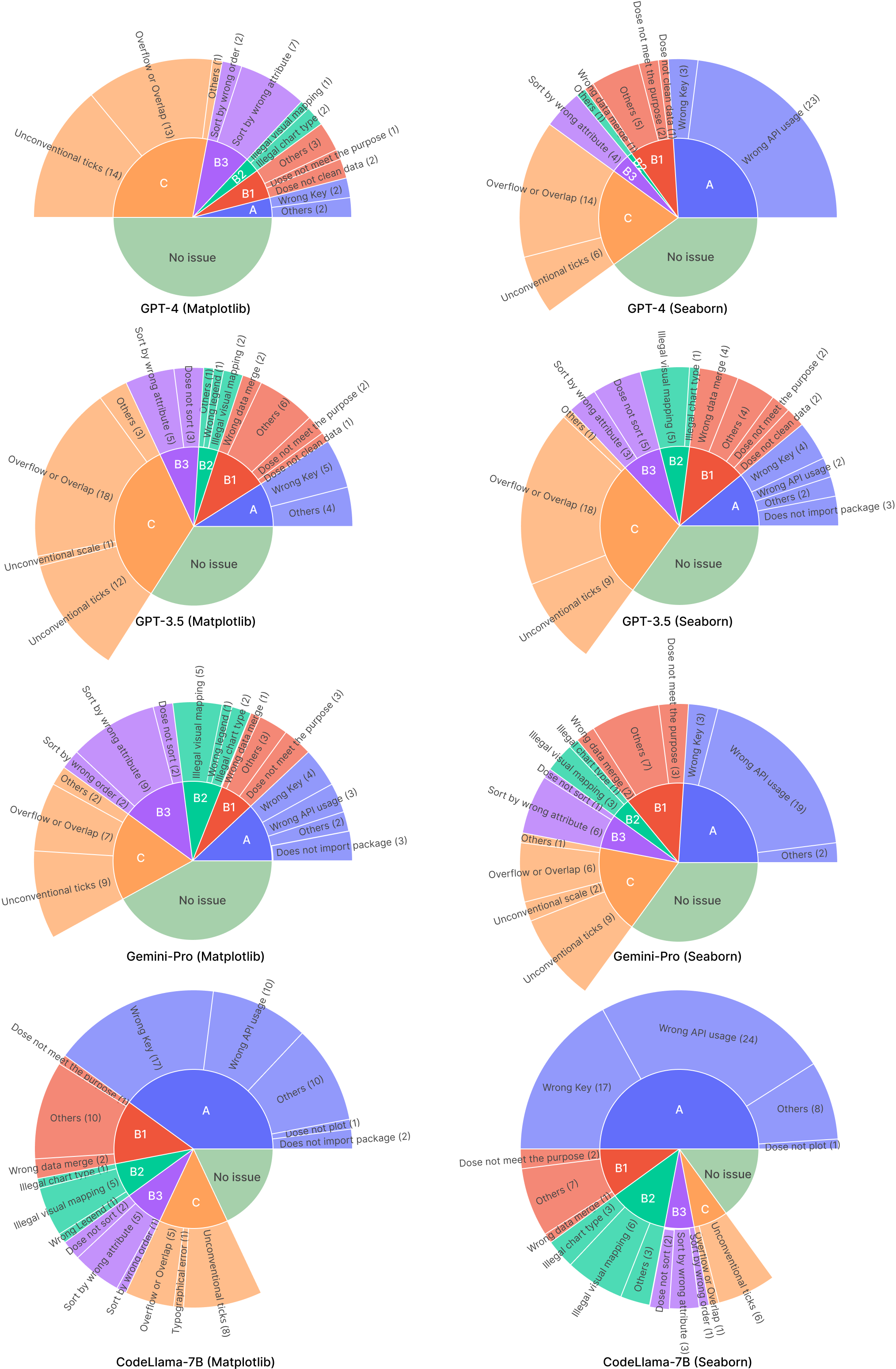}
    \vspace{0mm}
    \caption{\revise{Eight sunburst charts presenting a statistical analysis of common error causes. Each chart shows the number of chart errors from different causes for four models, using Matplotlib or Seaborn. Type A denotes invalid code errors. Types B1-B3 denote illegal data transformation, visualization transformation, and order errors, respectively. Type C denotes low readability issues.}}
    \label{fig:count}
    \vspace{2mm}
\end{figure*}

\revise{To gain insights into common errors, we manually reviewed visualizations generated by four models: GPT-4, GPT-3.5, Gemini-Pro, and CodeLlama-7B, using either Matplotlib or Seaborn. We randomly sampled 100 queries from our dataset, ensuring statistical representation. Our methodology involves leveraging feedback from evaluators, which includes issues such as crashes during code execution or detection of illegal data through data check. Through meticulous examination of generated code, chart images, and original queries, we identified the primary error for each visualization, acknowledging that each visualization may have multiple errors but focusing on the most significant one. \cref{fig:count} presents the results of this statistical analysis.}

For a more comprehensive understanding of error causes, we computed the error rate by each sub-check in the evaluation framework, which is the proportion of visualizations with errors divided by the total number of visualizations.
The reasons for errors are diverse. ~\autoref{tab:compare_detail} presents the error rate identified by different sub-checks across various models. ~\autoref{tab:compare_detail_approaches} displays the error rate identified by different sub-checks across different NL2VIS approaches for queries involving a single table.

\subsection{Evaluating Different Table Formats}
We compare the invalid rate, illegal rate, pass rate, readability score, and quality score across different table formats using Matplotlib library, as shown in~\autoref{tab:compare_data_format_detail}.

\input{tables/compare_detail}

\end{appendices}

%% file: tables/nvbench_issues.tex
\begin{table*}[bt!]
\small
\centering
\caption{Query issues in nvBench.}
\vspace{-2mm}
\renewcommand\arraystretch{1.2}
\setlength{\tabcolsep}{2mm}
\begin{tabular}{p{0.1\textwidth} p{0.59\textwidth} p{0.24\textwidth} }
\toprule
\textbf{Issue Type}    & \textbf{Example}      & \textbf{Explanation}      \\
\midrule
\multirow{4}{*}{\revise{\textbf{Irrationality}}} & What are the different first names and ages of the students who do have pets. Visualize by \error{pie chart}. \textit{(2562)} & \error{Inappropriate chart type}, age is not suitable for calculating proportions.\\
    & Scatter chart. what are the \error{faculty id} and the number of students each faculty has? & \error{Inappropriate attribute}, ID should not be treated as numerical. \\
\hline
\multirow{5}{*}{\revise{\textbf{Ambiguity}}}  & Show the number of male and female assistant professors. \textit{(21)}&\error{Unspecified chart type}, ground truth is a pie chart, not the optimal.\\ 
    & For those records from the products and each product's manufacturer, give me the comparison about the amount of \error{name} over the \error{name} , and group by attribute \error{name}, rank in desc by the names. \textit{(2204@x\_name@DESC)} & \error{Unclear attribute}, both product and manufacturer tables contain a column named ``name''.  \\
    & Show me about the distribution of date\_address\_to and the amount of date\_address\_to, bin date\_address\_to \error{by time} in a bar chart. \textit{(205)} & \error{Unspecified bin granularity}.  \\ 
    \hline
\multirow{4}{*}{\revise{\textbf{Duplication}}}  & Find the number of web accelerators used for each Operating system, \error{[empty]} / \error{list from low to high by the X-axis} / \error{display from high to low by the names please} / \error{I want to display by the y-axis in ascending} / \error{show from high to low by the total number}. \textit{(372, 372@x\_name@ASC, 372@x\_name@DESC, 372@y\_name@ASC, 372@y\_name@DESC)} & \error{Similar NL query} with different sorting requirement. \\
\hline
\multirow{4}{*}{\textbf{\revise{Erroneous lables}}} & \revise{Compare the prices of each manufacturer's most expensive product with a bar chart, and could you order x\-axis in descending order? \textit{(2203@x\_name@DESC)}} & \revise{Wrong x\_data Fujits (should be Fujitsu, the letter ``u'' was missing).} \\
& \revise{Show me about the distribution of date\_address\_to and the sum of monthly\_rental, and group by attribute other\_details and bin date\_address\_to by month in a bar chart. \textit{(194)}}& \revise{Incorrect y\_data 1297 (should be 1297.3186, the decimal was truncated).} \\
\bottomrule
\end{tabular}

\label{tab:benchmark-shortcomings}
\end{table*}

%% file: tables/compare_dataset.tex
\begin{table*}[htb]
\centering
\small
\caption{\revise{Comparison of existing visualization datasets.}}
\vspace{-2mm}
\begin{tabular}{c|cccccc}
\toprule
Name & Type & Queries & Databases & Visualization & Visualization types & Ground truth visualization \\
\midrule
VizNet~\cite{viznet} & Database only & $\times$  & 31M & $\times$  & $\times$ & $\times$ \\
Kim~\etal~\cite{kim2020answering} & Visualization question answering & 629 & 52 & 52 & 2 & $\times$\\
PlotQA~\cite{methani2020plotqa} & Visualization question answering & 28.9M & $\times$ & 224,377 & 3 & $\times$ \\
ChartQA~\cite{masry2022chartqa} & Visualization question answering & 32,719 & $\times$ & 21.9K & 3 & $\times$ \\
Chart-to-Text~\cite{kantharaj2022chart} & Chart summarization & $\times$ & $\times$ & 44,096 & 6 & $\times$\\
ChartSumm~\cite{rahman2022chartsumm} & Chart summarization & $\times$ & $\times$ & 84,363 & 3  & $\times$\\
Data2VIS~\cite{dibia2019data2vis} & Visualization recommendation & $\times$  & 11 & 4,300 & 6 & $\times$ \\
VizML~\cite{hu2019vizml} & Visualization recommendation & $\times$  & 119,815 & 119,815 & 3 & $\times$\\
Quda~\cite{fu2020quda} & NL2VIS & 14,035 & 36 & $\times$  & $\times$  & $\times$ \\
nvBench~\cite{luo2021synthesizing} & NL2VIS & 25,750 & 153 & 7,247 &7 & Contains inaccuracies  \\
NLV~\cite{srinivasan2021collecting} & NL2VIS & 893 & 3 & 30 & 10 & \checkmark \\
\hline
VisEval (ours) & NL2VIS & 2,524 & 146 & 1,150 & 7 & \checkmark \\

\bottomrule
\end{tabular}%
\vspace{-3mm}
\label{tab:compare_existing_dataset}
\end{table*}

%% file: tables/task.tex
\begin{table}[htb]
\centering
\small
\caption{\revise{Types of low-level tasks in queries.}}
\vspace{-3mm}
\begin{tabular}{c|cc}
\toprule
Analytical Task & Count & Proportion \\
\midrule
Compute Derived Value & 1915 & 75.87\% \\
Sort & 1230 & 48.73\% \\ 
Filter & 595 & 23.57\% \\
Retrieve Value & 575 & 22.78\% \\
Characterize Distribution & 441 & 17.47\% \\
Correlate & 208 & 8.24\% \\
Find Extreme & 117 & 4.64\% \\
Determine Range & 3 & 0.12\% \\
\bottomrule
\end{tabular}%
\vspace{-4mm}
\label{tab:task_types}
\end{table}

%% file: tables/compare_detail.tex
\begin{table*}[t]
\centering
\small
\caption{Comparison of the error rates identified by each sub-check module across different models.}
\vspace{-2mm}
\begin{tabular}{cc|cc|cccc|cc}
\toprule
\multirow{2}*{Model}& \multirow{2}*{Library} & \multicolumn{2}{c|}{Invalid} & \multicolumn{4}{c|}{Illegal} & \multicolumn{2}{c}{Low Readability}\\
& \begin{tabular}{@{}c@{}}\end{tabular}  & Execution & Surface-form & Deconstruction & Chart type & Data & Order & Layout & Scale\&Ticks\\ 
\midrule
CodeLlama-7B & \multirow{4}*{Matplotlib}  & {\cellcolor[HTML]{e4a9a9}} \color[HTML]{000000} 41.57\%  & {\cellcolor[HTML]{fefdfd}} \color[HTML]{000000} 1.38\%  & {\cellcolor[HTML]{ffffff}} \color[HTML]{000000} 0.29\%  & {\cellcolor[HTML]{fdf8f8}} \color[HTML]{000000} 3.54\%  & {\cellcolor[HTML]{f2d7d7}} \color[HTML]{000000} 19.60\%  & {\cellcolor[HTML]{f9eded}} \color[HTML]{000000} 8.90\%  & {\cellcolor[HTML]{fcf4f4}} \color[HTML]{000000} 5.43\%  & {\cellcolor[HTML]{faf0f0}} \color[HTML]{000000} 7.48\% \\Gemini-Pro &   & {\cellcolor[HTML]{f6e2e2}} \color[HTML]{000000} 14.35\%  & {\cellcolor[HTML]{ffffff}} \color[HTML]{000000} 0.00\%  & {\cellcolor[HTML]{ffffff}} \color[HTML]{000000} 0.05\%  & {\cellcolor[HTML]{fcf4f4}} \color[HTML]{000000} 5.14\%  & {\cellcolor[HTML]{f2d7d7}} \color[HTML]{000000} 19.64\%  & {\cellcolor[HTML]{f6e2e2}} \color[HTML]{000000} 14.36\%  & {\cellcolor[HTML]{fbf2f2}} \color[HTML]{000000} 6.34\%  & {\cellcolor[HTML]{f6e1e1}} \color[HTML]{000000} 14.71\% \\GPT-3.5 &   & {\cellcolor[HTML]{f9eded}} \color[HTML]{000000} 8.79\%  & {\cellcolor[HTML]{ffffff}} \color[HTML]{000000} 0.00\%  & {\cellcolor[HTML]{ffffff}} \color[HTML]{000000} 0.22\%  & {\cellcolor[HTML]{fdfafa}} \color[HTML]{000000} 2.42\%  & {\cellcolor[HTML]{f3d9d9}} \color[HTML]{000000} 18.67\%  & {\cellcolor[HTML]{f8eaea}} \color[HTML]{000000} 10.53\%  & {\cellcolor[HTML]{f2d5d5}} \color[HTML]{000000} 20.64\%  & {\cellcolor[HTML]{f1d3d3}} \color[HTML]{000000} 21.29\% \\GPT-4 &   & {\cellcolor[HTML]{fdf9f9}} \color[HTML]{000000} 3.29\%  & {\cellcolor[HTML]{ffffff}} \color[HTML]{000000} 0.00\%  & {\cellcolor[HTML]{ffffff}} \color[HTML]{000000} 0.28\%  & {\cellcolor[HTML]{fefbfb}} \color[HTML]{000000} 2.01\%  & {\cellcolor[HTML]{f8e8e8}} \color[HTML]{000000} 11.04\%  & {\cellcolor[HTML]{f9ecec}} \color[HTML]{000000} 9.61\%  & {\cellcolor[HTML]{f7e7e7}} \color[HTML]{000000} 12.02\%  & {\cellcolor[HTML]{eec9c9}} \color[HTML]{000000} 26.27\% \\
\hline
CodeLlama-7B &  \multirow{4}*{Seaborn} & {\cellcolor[HTML]{da8787}} \color[HTML]{000000} 57.85\%  & {\cellcolor[HTML]{fefdfd}} \color[HTML]{000000} 1.41\%  & {\cellcolor[HTML]{ffffff}} \color[HTML]{000000} 0.33\%  & {\cellcolor[HTML]{fcf7f7}} \color[HTML]{000000} 4.17\%  & {\cellcolor[HTML]{f5dfdf}} \color[HTML]{000000} 15.75\%  & {\cellcolor[HTML]{faf0f0}} \color[HTML]{000000} 7.70\%  & {\cellcolor[HTML]{fcf7f7}} \color[HTML]{000000} 3.96\%  & {\cellcolor[HTML]{fcf4f4}} \color[HTML]{000000} 5.17\% \\Gemini-Pro &   & {\cellcolor[HTML]{f2d4d4}} \color[HTML]{000000} 21.09\%  & {\cellcolor[HTML]{ffffff}} \color[HTML]{000000} 0.00\%  & {\cellcolor[HTML]{ffffff}} \color[HTML]{000000} 0.00\%  & {\cellcolor[HTML]{fcf6f6}} \color[HTML]{000000} 4.43\%  & {\cellcolor[HTML]{f4dddd}} \color[HTML]{000000} 16.44\%  & {\cellcolor[HTML]{f9ecec}} \color[HTML]{000000} 9.26\%  & {\cellcolor[HTML]{faf0f0}} \color[HTML]{000000} 7.07\%  & {\cellcolor[HTML]{f5e0e0}} \color[HTML]{000000} 15.13\% \\GPT-3.5 &   & {\cellcolor[HTML]{f9ecec}} \color[HTML]{000000} 9.21\%  & {\cellcolor[HTML]{ffffff}} \color[HTML]{000000} 0.00\%  & {\cellcolor[HTML]{ffffff}} \color[HTML]{000000} 0.18\%  & {\cellcolor[HTML]{fefbfb}} \color[HTML]{000000} 2.21\%  & {\cellcolor[HTML]{f4dcdc}} \color[HTML]{000000} 17.13\%  & {\cellcolor[HTML]{f7e4e4}} \color[HTML]{000000} 13.18\%  & {\cellcolor[HTML]{f2d7d7}} \color[HTML]{000000} 19.72\%  & {\cellcolor[HTML]{f3d9d9}} \color[HTML]{000000} 18.74\% \\GPT-4 &   & {\cellcolor[HTML]{efcaca}} \color[HTML]{000000} 25.41\%  & {\cellcolor[HTML]{ffffff}} \color[HTML]{000000} 0.00\%  & {\cellcolor[HTML]{ffffff}} \color[HTML]{000000} 0.13\%  & {\cellcolor[HTML]{fffefe}} \color[HTML]{000000} 0.77\%  & {\cellcolor[HTML]{f9ecec}} \color[HTML]{000000} 9.76\%  & {\cellcolor[HTML]{fbf3f3}} \color[HTML]{000000} 5.92\%  & {\cellcolor[HTML]{f6e2e2}} \color[HTML]{000000} 14.36\%  & {\cellcolor[HTML]{f6e1e1}} \color[HTML]{000000} 14.61\% \\
\bottomrule
\end{tabular}%

\label{tab:compare_detail}
\end{table*}

\begin{table*}[t]
\centering
\small
\caption{Comparison of the error rates identified by each sub-check module across different approaches for queries involving a single table. The evaluation results are obtained using the GPT-3.5 model.}
\vspace{-2mm}
\begin{tabular}{cc|cc|cccc|cc}
\toprule
\multirow{2}*{Model}& \multirow{2}*{Library} & \multicolumn{2}{c|}{Invalid} & \multicolumn{4}{c|}{Illegal} & \multicolumn{2}{c}{Low Readability}\\
& \begin{tabular}{@{}c@{}}\end{tabular}  & Execution & Surface-form & Deconstruction & Chart Type & Data & Order & Layout & Scale\&Ticks\\ 
\midrule
CoML4VIS & \multirow{3}*{Matplotlib} & {\cellcolor[HTML]{fcf5f5}} \color[HTML]{000000} 4.95\%  & {\cellcolor[HTML]{ffffff}} \color[HTML]{000000} 0.00\%  & {\cellcolor[HTML]{ffffff}} \color[HTML]{000000} 0.37\%  & {\cellcolor[HTML]{fdf9f9}} \color[HTML]{000000} 3.07\%  & {\cellcolor[HTML]{f4dcdc}} \color[HTML]{000000} 17.08\%  & {\cellcolor[HTML]{f9ebeb}} \color[HTML]{000000} 10.04\%  & {\cellcolor[HTML]{f1d1d1}} \color[HTML]{000000} 22.27\%  & {\cellcolor[HTML]{f1d2d2}} \color[HTML]{000000} 22.09\% \\LIDA &   & {\cellcolor[HTML]{fbf2f2}} \color[HTML]{000000} 6.59\%  & {\cellcolor[HTML]{ffffff}} \color[HTML]{000000} 0.00\%  & {\cellcolor[HTML]{fefdfd}} \color[HTML]{000000} 1.06\%  & {\cellcolor[HTML]{fefbfb}} \color[HTML]{000000} 2.31\%  & {\cellcolor[HTML]{f6e4e4}} \color[HTML]{000000} 13.50\%  & {\cellcolor[HTML]{f8e8e8}} \color[HTML]{000000} 11.07\%  & {\cellcolor[HTML]{de9595}} \color[HTML]{000000} 51.42\%  & {\cellcolor[HTML]{f0cfcf}} \color[HTML]{000000} 23.49\% \\Chat2vis &   & {\cellcolor[HTML]{fcf5f5}} \color[HTML]{000000} 4.92\%  & {\cellcolor[HTML]{ffffff}} \color[HTML]{000000} 0.00\%  & {\cellcolor[HTML]{fffefe}} \color[HTML]{000000} 0.41\%  & {\cellcolor[HTML]{fdf9f9}} \color[HTML]{000000} 2.85\%  & {\cellcolor[HTML]{f6e4e4}} \color[HTML]{000000} 13.57\%  & {\cellcolor[HTML]{f8e8e8}} \color[HTML]{000000} 11.10\%  & {\cellcolor[HTML]{e3a6a6}} \color[HTML]{000000} 43.13\%  & {\cellcolor[HTML]{f0cfcf}} \color[HTML]{000000} 23.12\% \\
\hline
CoML4VIS &  \multirow{3}*{Seaborn} & {\cellcolor[HTML]{fbf1f1}} \color[HTML]{000000} 6.95\%  & {\cellcolor[HTML]{ffffff}} \color[HTML]{000000} 0.00\%  & {\cellcolor[HTML]{ffffff}} \color[HTML]{000000} 0.22\%  & {\cellcolor[HTML]{fdfafa}} \color[HTML]{000000} 2.73\%  & {\cellcolor[HTML]{f5dede}} \color[HTML]{000000} 16.35\%  & {\cellcolor[HTML]{f6e1e1}} \color[HTML]{000000} 14.61\%  & {\cellcolor[HTML]{f3d9d9}} \color[HTML]{000000} 18.73\%  & {\cellcolor[HTML]{f3dada}} \color[HTML]{000000} 18.08\% \\LIDA &   & {\cellcolor[HTML]{f7e5e5}} \color[HTML]{000000} 12.69\%  & {\cellcolor[HTML]{ffffff}} \color[HTML]{000000} 0.00\%  & {\cellcolor[HTML]{fefdfd}} \color[HTML]{000000} 0.86\%  & {\cellcolor[HTML]{fdf9f9}} \color[HTML]{000000} 3.45\%  & {\cellcolor[HTML]{f4dddd}} \color[HTML]{000000} 16.43\%  & {\cellcolor[HTML]{f4dbdb}} \color[HTML]{000000} 17.60\%  & {\cellcolor[HTML]{f0cece}} \color[HTML]{000000} 23.94\%  & {\cellcolor[HTML]{f6e2e2}} \color[HTML]{000000} 14.42\% \\Chat2vis &   & {\cellcolor[HTML]{fdf8f8}} \color[HTML]{000000} 3.65\%  & {\cellcolor[HTML]{ffffff}} \color[HTML]{000000} 0.00\%  & {\cellcolor[HTML]{ffffff}} \color[HTML]{000000} 0.25\%  & {\cellcolor[HTML]{fdf9f9}} \color[HTML]{000000} 3.37\%  & {\cellcolor[HTML]{f6e1e1}} \color[HTML]{000000} 14.75\%  & {\cellcolor[HTML]{f5e0e0}} \color[HTML]{000000} 14.99\%  & {\cellcolor[HTML]{e4a8a8}} \color[HTML]{000000} 42.38\%  & {\cellcolor[HTML]{f2d5d5}} \color[HTML]{000000} 20.39\% \\
\bottomrule
\end{tabular}%
\label{tab:compare_detail_approaches}
\end{table*}

\begin{table*}[t]
\centering
\small
\caption{Comparison of the invalid rate, illegal rate, pass rate, readability score, and quality score across different table formats using Matplotlib.}
\vspace{-2mm}
\begin{tabular}{cc|cc|c|c|c}
\toprule
Model & Table Format &  \begin{tabular}{@{}c@{}}Invalid \\ Rate \end{tabular}  & \begin{tabular}{@{}c@{}}Illegal \\ Rate\end{tabular} & \begin{tabular}{@{}c@{}}Pass \\ Rate \end{tabular} & \begin{tabular} {@{}c@{}}Readability \\ Score \end{tabular}   & \begin{tabular}{@{}c@{}}Quality \\ Score \end{tabular} \\
\midrule
Gemini-Pro &  \multirow{3}*{CoML} & {\cellcolor[HTML]{f6e2e2}} \color[HTML]{000000} 14.35\%  & {\cellcolor[HTML]{e9b9b9}} \color[HTML]{000000} 34.06\%  & {\cellcolor[HTML]{93c799}} \color[HTML]{000000} 51.59\%  & {\cellcolor[HTML]{429d4d}} \color[HTML]{ffffff} 3.95  & {\cellcolor[HTML]{7dbb84}} \color[HTML]{000000} 2.06 \\GPT-3.5 &   & {\cellcolor[HTML]{f9eded}} \color[HTML]{000000} 8.79\%  & {\cellcolor[HTML]{ecc2c2}} \color[HTML]{000000} 29.42\%  & {\cellcolor[HTML]{7ebb85}} \color[HTML]{000000} 61.79\%  & {\cellcolor[HTML]{f7fbf7}} \color[HTML]{000000} 3.52  & {\cellcolor[HTML]{70b478}} \color[HTML]{000000} 2.21 \\GPT-4 &   & {\cellcolor[HTML]{fdf9f9}} \color[HTML]{000000} 3.29\%  & {\cellcolor[HTML]{f1d3d3}} \color[HTML]{000000} 21.44\%  & {\cellcolor[HTML]{62ad6b}} \color[HTML]{ffffff} \bfseries 75.27\%  & {\cellcolor[HTML]{82be89}} \color[HTML]{000000} 3.80  & {\cellcolor[HTML]{379742}} \color[HTML]{ffffff} \bfseries 2.89 \\
\hline
Gemini-Pro & \multirow{3}*{LIDA} & {\cellcolor[HTML]{f4dddd}} \color[HTML]{000000} 16.58\%  & {\cellcolor[HTML]{efcbcb}} \color[HTML]{000000} 25.03\%  & {\cellcolor[HTML]{85bf8c}} \color[HTML]{000000} 58.39\%  & {\cellcolor[HTML]{68b071}} \color[HTML]{000000} 3.86  & {\cellcolor[HTML]{6bb273}} \color[HTML]{000000} 2.27 \\GPT-3.5 &   & {\cellcolor[HTML]{faefef}} \color[HTML]{000000} 8.20\%  & {\cellcolor[HTML]{ecc3c3}} \color[HTML]{000000} 29.28\%  & {\cellcolor[HTML]{7cbb83}} \color[HTML]{000000} 62.52\%  & {\cellcolor[HTML]{ffffff}} \color[HTML]{000000} 3.47  & {\cellcolor[HTML]{71b579}} \color[HTML]{000000} 2.19 \\GPT-4 &   & {\cellcolor[HTML]{fcf7f7}} \color[HTML]{000000} 4.08\%  & {\cellcolor[HTML]{f1d3d3}} \color[HTML]{000000} 21.26\%  & {\cellcolor[HTML]{62ad6b}} \color[HTML]{000000} 74.66\%  & {\cellcolor[HTML]{8ec494}} \color[HTML]{000000} 3.77  & {\cellcolor[HTML]{3b9946}} \color[HTML]{FFFFFF} 2.84 \\
\hline
Gemini-Pro &  \multirow{3}*{Chat2vis} & {\cellcolor[HTML]{f4dbdb}} \color[HTML]{000000} 17.68\%  & {\cellcolor[HTML]{f0cece}} \color[HTML]{000000} 23.99\%  & {\cellcolor[HTML]{85bf8c}} \color[HTML]{000000} 58.33\%  & {\cellcolor[HTML]{58a861}} \color[HTML]{000000} 3.90  & {\cellcolor[HTML]{69b172}} \color[HTML]{000000} 2.29 \\GPT-3.5 &   & {\cellcolor[HTML]{faefef}} \color[HTML]{000000} 8.01\%  & {\cellcolor[HTML]{edc6c6}} \color[HTML]{000000} 27.98\%  & {\cellcolor[HTML]{79b981}} \color[HTML]{000000} 64.01\%  & {\cellcolor[HTML]{ffffff}} \color[HTML]{000000} 3.42  & {\cellcolor[HTML]{6fb477}} \color[HTML]{000000} 2.22 \\GPT-4 &   & {\cellcolor[HTML]{fcf5f5}} \color[HTML]{000000} 4.74\%  & {\cellcolor[HTML]{f0cfcf}} \color[HTML]{000000} 23.35\%  & {\cellcolor[HTML]{68b071}} \color[HTML]{000000} 71.91\%  & {\cellcolor[HTML]{abd3b0}} \color[HTML]{000000} 3.70  & {\cellcolor[HTML]{479f52}} \color[HTML]{FFFFFF} 2.69 \\
\bottomrule
\end{tabular}%
\label{tab:compare_data_format_detail}
\end{table*}